\documentclass[11pt]{article}\include{ddefs}
\usepackage{graphics,pifont,array}
\usepackage{graphicx}
\usepackage{endnotes}
\let\footnote=\endnote

\let\footnote=\endnote

\begin{document}

\begin{titlepage}

	\begin{center}

		{\Large{        \mbox{   }                          \\
				\mbox{   }                          \\
				\mbox{   }                          \\
				\mbox{   }                          \\
				\mbox{   }                          \\
				\mbox{   }                          \\
				\mbox{   }                          \\
				\mbox{   }                          \\
				\mbox{   }                          \\
				\mbox{   }                          \\
 {\textbf{DIELECTRIC SLAB REFLECTION/TRANSMISSION   \\
 	AS A SELF-CONSISTENT RADIATION PHENOMENON}}	    \\
				\mbox{    }                         \\
				\mbox{    }                         \\
				J. A. Grzesik                       \\
				Allwave Corporation                 \\
				3860 Del Amo Boulevard              \\
				Suite 404                           \\
				Torrance, CA 90503                  \\
				\mbox{    }                         \\
				(818) 749-3602                      \\ 
				jan.grzesik@hotmail.com             \\
				\mbox{     }                        \\
				\mbox{     }                        \\  
				\today                                   }  }
		
	\end{center}

\end{titlepage}

\setcounter{page}{2}

\pagenumbering{roman}
\setcounter{page}{2}
\vspace*{1.0in}


\begin{abstract}
	We revisit the standard electromagnetic problem wherein wave propagation within a uniform,	
lossless dielectric is interrupted by a dissipative slab of finite thickness.  While such a problem
is easily solved on the basis of interface field continuity, we proceed to treat it here
under the viewpoint of radiative self-consistency, with effective current
sources resident only within the slab interior and gauged by ohmic/polarization parameter
comparisons against those of the reference, exterior medium.  Radiative self-consistency finds
its natural expression as an integral equation over the slab interior field which, once solved, permits
a direct, fully constructive buildup, both up and down, of the reflected/transmitted field
contributions, without any need for ascertaining such quantities implicitly via the enforcement of
boundary conditions.  The persistent cadence of solution steps in such integral-equation problems
asserts itself here, too, in the sense that it leads, first, to an exact cancellation, left and right,
of that interior, unknown field, and second, that it brings in still other contributions of a
reference medium variety, of which it is required that they, and only they, balance the incoming
excitation.  Balancing of this latter sort provides indeed the linear conditions for slab field
determination.  The two-step solution pattern thus described may be regarded as a manifestation
at some remove of Ewald-Oseen extinction, even though the analytic framework now on view differs
fundamentally from proofs elsewhere available.  We go on to solve the several balancing equations by
direct, vector manipulation avoiding all recourse to determinants, and then offer a partial confirmation
by exhibiting a canonical, boundary value counterpart in the special case of perpendicular incidence.
Following all of this, in an appendix, we allow one of the enclosing half spaces to differ from
that wherein the excitation has been launched and which continues to serve as the reference medium.
Effective currents are now found not only within the slab proper, but also throughout an entire
half space, necessitating a suitable generalization of the underlying integral equation, and a
provision, during its solution, of cross-talk, both up and down, between slab and the half space
now contributing as a radiation source.  We provide in this appendix a fairly accelerated presentation
of all these generalized features, but stop a shade short of an explicit field solution by reason
of a galloping algebraic inflation.  All logical details are however displayed in plain view.
The integral equation radiative self-consistency method is, to our way of thinking, physically
far more satisfying than the prevailing method of scattered fields guessed as to their structure
and then fixed by boundary conditions.  Its analytic themes, moreover, are far, far more elegant.
\end{abstract}


\parindent=0.5in

\newpage

\pagenumbering{arabic}

\pagestyle{myheadings}

\setlength{\parindent}{0pt}

\pagestyle{plain}

\parindent=0.5in

\newpage
\mbox{   }

\pagestyle{myheadings}

\markright{J. A. Grzesik \\ dielectric slab reflection/transmission as a self-consistent radiation phenomenon}

\section{Introduction}
      We wish here to revisit from a radically fresh viewpoint the standard problem of plane-wave electromagnetic     
reflection from, and transmission across, a lossy dielectric slab.  Our motivation for doing so is rooted in
methodological, pedagogical, and, above all else, in physical considerations.  Indeed, we shall show that both the
so-called reflected and transmitted fields arise as the cumulative radiation from the self-consistent ohmic/dielectric
polarization currents excited within the slab interior.  In particular, under this perspective, boundary matching
conditions of the standard sort will never need to be invoked.  This, to our way of thinking at least, provides a
physically much more satisfying picture of, say, the reflected wave, which, in normal, glib parlance, embodies an
implied behavior resembling some type of interface bounce, and never mind that the ensuing boundary matching program
does in fact yield correct results.  A bounce there indeed is, but in the developments that follow we, so to speak,
dissect its radiative anatomy, its radiative genesis.

     Pedagogy and methodology are of course inextricably intertwined, and here both benefit in conviction by being
allowed to unfold in a setting whose resolution is available by other means.  The problem that we envision involves
a lossy dielectric slab of thickness $a,$ permittivity $\epsilon_{2},$ conductivity $\sigma_{2},$ sandwiched between
two identical,
nondissipative dielectric half spaces, reference permittivity $\epsilon_{1}\neq\epsilon_{2},$ reference conductivity
$\sigma_{1}=0.$\footnote{In an appendix we set out at some length the type of analytic complications which ensue should
dissimilar exterior half spaces be admitted into the discussion.  Although not insurmountable, such complications are
best avoided in an initial, proof-of-principle discussion such as that which now confronts us.  Moreover, so as not to
unduly complicate our slab problem, we further insist upon strict magnetic uniformity, with a common magnetic permeability
$\mu_{2}=\mu_{1}=\mu.$}  A plane wave propagates downward from above, traversing the slab, and continuing into the half
space below.  But, because the slab differs from the reference medium, a current density
$(\sigma_{2}-i\omega\{\epsilon_{2}-\epsilon_{1}\}){\bf{  {E}}}_{{\rm{tot}}},$ with ${\bf{  {E}}}_{{\rm{tot}}}$ being
the local self-consistent, {\em{total}} electric field, incident plus radiated,
is set up throughout its interior.\footnote{A simple harmonic time dependence $\exp(-i\omega t),$ $\omega$ being of
either sign, is assumed
throughout and simply removed from all equations as a non-participating factor.}  Radiation from these currents is then
superimposed upon the primary in such a way as to establish that self-consistent field
${\bf{  {E}}}_{{\rm{tot}}}$ on both interior and exterior, and thus to modify
as needed the plane wave vector amplitude in the half space below, and, of course, to assemble the plane-wave echo, due to
slab presence, in the half space of wave origin above.  When duly asserted within the slab interior, this statement
of radiative self-consistency takes the form of an integral equation, easily derived and recorded as Eq. (9) below.

    Once that integral equation is firmly before us, we shall move quickly to set down the interior slab field with
up/down vector amplitudes ${\bf{  {E}}}_{{\rm{slab}}}^{\,\pm},$ momentarily unknown, and to delineate the {\em{bona fide}}
plane-wave aspect of the fields radiated into both enclosing half spaces.  When framing the structure of the
self-consistent interior field we shall naturally incorporate as much {\em{a priori}} information as the underlying Maxwell
differential equations permit, but will of course stop short of reciting such equations anew.  Following this, we
shall turn to the main task of actually solving Eq. (9) for	
${\bf{  {E}}}_{{\rm{slab}}}^{\,\pm}$ by methods which, while still intricate, are free from all allusion to boundary matching
and the machinery of determinants.  This work, in particular, will proceed without any orientational constraint upon the
incoming plane wave vector amplitude ${\bf{  {E}}}_{{\rm{inc}}},$ apart, of course, from its obligatory orthogonality to the
sense of propagation ${\rm{\bf{\hat{n}}}},$ ${\rm{\bf{\hat{n}}}}{\mbox{\boldmath$\,\cdot\,$}}
{\bf{  {E}}}_{{\rm{inc}}}=0.$  A concluding section will then specialize our formulae
so as to recover in every detail the classical Fresnel results, at least in the simplest, default situation of
perpendicular incidence.
\newpage
\mbox{   }
\section{Problem setup}
     Origin ${\bf{0}}$ of a right-handed Cartesian co{\"{o}}rdinate system is taken to lie on the upper, wave incidence
slab side, with unit vertical vector ${\rm{\bf{\hat{e}}}}_{z}$ pointing upward, against the sense of arrival, and both
$x$ and $y$ axes thus contained in the material interface.\footnote{The corresponding basis vectors in the $x$-$y$ plane are
${\rm{\bf{\hat{e}}}}_{x}$ and  ${\rm{\bf{\hat{e}}}}_{y},$ with vector
${\mbox{\boldmath$\rho$}}=x\,{\rm{\bf{\hat{e}}}}_{x}+y\,{\rm{\bf{\hat{e}}}}_{y}$
designating horizontal displacement from a vertical axis passing through origin ${\bf{0}}.$  A full three-dimensional displacement ${\rm{\bf{r}}}$ from
${\bf{0}}$ is then rendered as ${\rm{\bf{r}}}={\mbox{\boldmath$\rho$}}+z\,{\rm{\bf{\hat{e}}}}_{z}.$}
The incident plane wave is considered to propagate along ${\rm{\bf{\hat{n}}}}={\rm{\bf{\hat{n}}}}${\scriptsize{$_{\|}$}}$+\,{\hat{n}}_{z}{\rm{\bf{\hat{e}}}}_{z},$ with its
vertical component ${\hat{n}}_{z}$ negative, ${\hat{n}}_{z}<0,$ and ${\rm{\bf{\hat{n}}}}${\scriptsize{$_{\|}$}}
being its $x$-$y$ projection.

     The impinging wave has a propagation constant $k_{1}=\omega\sqrt{\rule{0mm}{3mm}\epsilon_{1}\mu\,}$ and
a spatial behavior governed by a phase factor
$\exp\left(\rule{0mm}{3mm}k_{1}{\rm{\bf{\hat{n}}}}{\mbox{\boldmath$\cdot$}}{\rm{\bf{r}}}\right),$ whose horizontal
evolution in accordance with
$\exp\left(\rule{0mm}{1.5mm}ik_{1}{\rm{\bf{\hat{n}}}}_{\|}
{\mbox{\boldmath$\cdot$}}{\mbox{\boldmath$\rho$}}\right)$ is imprinted upon all participating field
categories, both within and exterior to the slab.  Such commonality responds to the patent need to maintain a
compatible horizontal phase progression everywhere, and is in fact quite spontaneously reaffirmed by our 
integral, radiative framework.

    And so, if, for $-a<z<0,$ we write the total self-consistent slab field as
\begin{eqnarray}
{\bf{  {E}}}_{{\rm{tot}}}({\bf{r}}) & = & \sum_{\pm}{\bf{  {E}}}_{{\rm{slab}}}^{\pm}
  \exp\left(\rule{0mm}{1.5mm}ik_{1}{\rm{\bf{\hat{n}}}}_{\|}
                    {\mbox{\boldmath$\cdot$}}{\mbox{\boldmath$\rho$}}+ p_{\pm}z\right),
\end{eqnarray}
then a demand that it conform to an underlying Helmholtz equation requires that vertical propagation parameters $p_{\pm}$ obey
\begin{equation}
p_{\pm}^{2}
-k_{1}^{2}{\rm{\bf{\hat{n}}}}_{\|}{\mbox{\boldmath$\cdot\,$}}{\rm{\bf{\hat{n}}}}_{\|}=
-\omega^{2}\mu\epsilon_{2}\left(\rule{0mm}{4mm}1+i\sigma_{2}/\omega\epsilon_{2}\right)\,.
\end{equation}
On further setting ${\rm{\bf{\hat{n}}}}{\mbox{\boldmath$\cdot\,$}}{\rm{\bf{\hat{e}}}}_{z}={\hat{n}}_{z}=-\cos\vartheta,$ with
incidence angle $\vartheta$ bounded from above and below as $-\pi/2<\vartheta<\pi/2,$ we
find that ${\rm{\bf{\hat{n}}}}_{\|}{\mbox{\boldmath$\cdot\,$}}{\rm{\bf{\hat{n}}}}_{\|}=\sin^{2}\vartheta$
and hence
\begin{eqnarray}
p_{\pm} &  = & \pm  k_{1}\sqrt{\rule{0mm}{4mm}\sin^{2}\vartheta-\left(\rule{0mm}{4mm}\epsilon_{2}/\epsilon_{1}\right)
	\!\left(\rule{0mm}{4mm}1+i\sigma_{2}/\omega\epsilon_{2}\right)\,} \nonumber  \\
& = & \pm\frac{k_{1}}{\sqrt{2\,}} \left\{\rule{0mm}{8mm}
\sqrt{\sqrt{\rule{0mm}{4mm}\left(\rule{0mm}{4mm}\sin^{2}\vartheta-\epsilon_{2}/\epsilon_{1}\right)^{2}+
	\left(\rule{0mm}{4mm}\sigma_{2}/\omega\epsilon_{1}\right)^{2}\,}+\sin^{2}\vartheta-\epsilon_{2}/\epsilon_{1}\,}  \right.  \\
&    &  \rule{-0.1cm}{0mm} \left.\rule{0mm}{8mm} -i\frac{\omega}{|\,\omega\,|}
\sqrt{\sqrt{\rule{0mm}{4mm}\left(\rule{0mm}{4mm}\sin^{2}\vartheta-\epsilon_{2}/\epsilon_{1}\right)^{2}+
			\left(\rule{0mm}{4mm}\sigma_{2}/\omega\epsilon_{1}\right)^{2}\,}-\sin^{2}\vartheta+\epsilon_{2}/\epsilon_{1}\,}  \;\right\},\nonumber			
\end{eqnarray}
which latter pays due deference to the sign of angular frequency $\omega.$
It then follows that index $+$ in (1) selects a wave component whose amplitude diminishes with penetration into the
slab, whereas the complementary index $-$ accompanies amplitude growth.  Observe that expression (3) accepts as valid
the ordering $0<\epsilon_{2}<\epsilon_{1},$ so that our slab could in fact represent a thinning, a soft gap, in some sense a
dilution, an evacuation of material in comparison with its two abutting half spaces.  The half-angle trigonometric formulae
which underlie the nested roots in (3) must track their angular arguments with great care when arriving at the stated formula.  
\newpage
\mbox{    }
\newline

     Field (1) must in addition be solenoidal, with vanishing divergence, circumstance which requires that
vector amplitudes ${\bf{  {E}}}_{{\rm{slab}}}^{\,\pm}$ submit to the null dot products
\begin{equation}
{\bf{   {P}}}^{\,\pm}{\mbox{\boldmath$\cdot\,$}}{\bf{  {E}}}_{{\rm{slab}}}^{\,\pm}=0\,,
\end{equation}
with
\begin{equation}
{\bf{   {P}}}^{\,\pm}=ik_{1}{\rm{\bf{\hat{n}}}}_{\|}+p_{\pm}{\rm{\bf{\hat{e}}}}_{z}\,.
\end{equation}
\section{Basic Integral Equation}
     We advocate the elementary viewpoint that the $\epsilon_{1}$ medium of the enclosing half spaces provides
a standard wave propagation environment disrupted only by the ohmic/excess polarization current density
$(\sigma_{2}-i\omega\{\epsilon_{2}-\epsilon_{1}\}){\bf{  {E}}}_{{\rm{tot}}}({\bf{r}})$ encountered across the vertical
slot $0>z>-a.$ Such currents then radiate {\em{everywhere}} a scattered magnetic field
\begin{equation}
{\bf{  {B}}}_{{\rm{scatt}}}({\bf{r}})=\frac{\mu(\sigma_{2}-i\omega\{\epsilon_{2}-\epsilon_{1}\})}{4\pi}
\sum_{\pm} 
           {\mbox{\boldmath$\nabla\times$}}\left\{\rule{0mm}{7mm}{\bf{  {E}}}_{{\rm{slab}}}^{\pm}
     \int_{{\rm{slab}}}\frac{e^{ik_{1}|{\bf{r}}-{\bf{r'}}|}}{|{\bf{r}}-{\bf{r'}}|}     
     	\exp\left(\rule{0mm}{1.5mm}ik_{1}{\rm{\bf{\hat{n}}}}_{\|}
     	{\mbox{\boldmath$\cdot$}}{\mbox{\boldmath${\tiny{\rho'}}$}}+ p_{\pm}z'\right)dx'dy'dz'\right\} 
\end{equation}
which combines linearly with that incident,
${\bf{  {B}}}_{{\rm{inc}}}({\bf{r}})=(k_{1}/\omega){\rm{\bf{\hat{n}}}}{\mbox{\boldmath$\,\times$}}{\bf{  {E}}}_{{\rm{inc}}}
\exp\left(\rule{0mm}{4mm}ik_{1}{\rm{\bf{\hat{n}}}}{\mbox{\boldmath$\cdot$}}{\bf{r}}\right),$ to yield a total
\begin{eqnarray}
\lefteqn{
	{\bf{  {B}}}_{{\rm{tot}}}({\bf{r}})  =
	\frac{k_{1}}{\omega}
	{\rm{\bf{\hat{n}}}}{\mbox{\boldmath$\,\times$}}{\bf{  {E}}}_{{\rm{inc}}}
	e^{ik_{1}{\rm{\bf{\hat{n}}}}{\mbox{\boldmath$\cdot$}}{\bf{r}}} } \nonumber \\
& & +\frac{\mu(\sigma_{2}-i\omega\{\epsilon_{2}-\epsilon_{1}\})}{4\pi}
\sum_{\pm}{\mbox{\boldmath$\,\nabla\times$}}\left\{\rule{0mm}{7mm}	{\bf{  {E}}}_{{\rm{slab}}}^{\pm}
\int_{{\rm{slab}}}\frac{e^{ik_{1}|{\bf{r}}-{\bf{r'}}|}}{|{\bf{r}}-{\bf{r'}}|}     
\exp\left(\rule{0mm}{1.5mm}ik_{1}{\rm{\bf{\hat{n}}}}_{\|}
{\mbox{\boldmath$\cdot$}}{\mbox{\boldmath$\rho'$}}+ p_{\pm}z'\right)dx'dy'dz'\right\}.
\end{eqnarray}
If we next define an effective dielectric permeability ${\hat{\epsilon}}(z)$ by setting
\begin{equation}
{\hat{\epsilon}}(z)=\left\{\begin{array}{lcl}
	        \epsilon_{1} & ; & z>0 \\
	        \epsilon_{2}+i\sigma_{2}/\omega & ; & 0>z>-a \\
	        \epsilon_{1} & ; & -a>z
	        \end{array} \
	        \right. \,,
\end{equation}
we find that the total electric field ${\bf{  {E}}}_{{\rm{tot}}}({\bf{r}})$ accompanying (7) is gotten
as
\begin{eqnarray}
\lefteqn{
{\hat{\epsilon}}(z){\bf{  {E}}}_{{\rm{tot}}}({\bf{r}})  =  
       \epsilon_{1}{\bf{  {E}}}_{{\rm{inc}}}e^{ik_{1}{\rm{\bf{\hat{n}}}}{\mbox{\boldmath$\cdot$}}{\bf{r}}} } \nonumber \\
 & &\rule{-7mm}{0mm} +\frac{i(\sigma_{2}-i\omega\{\epsilon_{2}-\epsilon_{1}\})}{4\pi\omega}
 \sum_{\pm}{\mbox{\boldmath$\nabla\times$}}\left[\rule{0mm}{7mm} 
 {\mbox{\boldmath$\,\nabla\times$}}\left\{\rule{0mm}{7mm}	{\bf{  {E}}}_{{\rm{slab}}}^{\pm}
 \int_{{\rm{slab}}}\frac{e^{ik_{1}|{\bf{r}}-{\bf{r'}}|}}{|{\bf{r}}-{\bf{r'}}|}     
 \exp\left(\rule{0mm}{1.5mm}ik_{1}{\rm{\bf{\hat{n}}}}_{\|}
 {\mbox{\boldmath$\cdot$}}{\mbox{\boldmath$\rho'$}}+ p_{\pm}z'\right)dx'dy'dz'\right\}\,\right] \rule{0mm}{7mm}  \,.
\end{eqnarray}
Equations (7) and (9) hold everywhere.  Evaluated respectively for $z>0$ and $z<-a,$ they provide a direct buildup of both
reflected and transmitted fields, whereas enforcement across the slab interior $0>z>-a$ determines the key interior
amplitudes ${\bf{  {E}}}_{{\rm{slab}}}^{\pm}.$  Total field buildup as a sum of incident plus scattered contributions,
while assuredly not a novel concept, receives a concise statement, akin to (9), in [{\bf{1}}] and [{\bf{2}}], and doubtless
elsewhere.
\newpage
\mbox{   }
\section{Preliminary calculations}
     The slab integral on the right in (9) is easily performed with the aid of well known results concerning Bessel function
$J_{0}.$  Thus a temporary shift of the horizontal origin to ${\mbox{\boldmath$\rho$}},$ followed by quadrature across the full,
$0$ to $2\pi$ azimuthal range, gives\footnote{$\rule{0mm}{1.8cm}$We call attention to the fact that ${\hat{n}}_{z},$
here and in all that follows,
is negative.  That the third line of Eq. (10) duly accommodates this attribute is guaranteed by meticulous checks against
the Bessel quadrature formulae found in [{\bf{3}}].  It is further reaffirmed, and much more directly so, by the elementary
quadrature which is made available on setting $\vartheta=0.$  The very convergence of this latter quadrature, and its succinct,
convenient outcome, are predicated as always on $k_{1}$ having a small residual dissipative part, regardless of the sign of
angular frequency $\omega.$} 
\begin{eqnarray}
\int_{{\rm{slab}}}\frac{e^{ik_{1}|{\bf{r}}-{\bf{r'}}|}}{|{\bf{r}}-{\bf{r'}}|}     
	\exp\left(ik_{1}{\rm{\bf{\hat{n}}}}_{\|}
	{\mbox{\boldmath$\cdot$}}{\mbox{\boldmath$\rho'$}}+ p_{\pm}z'\right)dx'dy'dz' & = &     \nonumber  \\
  &  & \rule{-6.6cm}{0mm}=\,2\pi \exp\left(ik_{1}{\rm{\bf{\hat{n}}}}_{\|}
{\mbox{\boldmath$\cdot$}}{\mbox{\boldmath$\rho$}}\right)\int_{-a}^{\,0}dz'\exp\left(\rule{0mm}{3.5mm}p_{\pm}z'\right)
\int_{0}^{\,\infty}
\frac{e^{ik_{1}\sqrt{\rule{0mm}{3mm}\rho\,'^{\,2}+(z-z')^{2}\,}}}{\sqrt{\rule{0mm}{3mm}\rho\,'^{\,2}+(z-z')^{2}\,}}\,  
\rho'J_{0}(k_{1}\rho'\sin\vartheta)d\rho'   \nonumber   \\
 &   & \rule{-6.6cm}{0mm}=\, \frac{2\pi}{ik_{1}{\hat{n}}_{z}}\exp\left(ik_{1}{\rm{\bf{\hat{n}}}}_{\|}
 {\mbox{\boldmath$\cdot$}}{\mbox{\boldmath$\rho$}}\right)\int_{-a}^{\,0}
 \exp\left(\rule{0mm}{3.5mm}p_{\pm}z'-ik_{1}{\hat{n}}_{z}|z-z'|\right)dz'\,.   
\end{eqnarray}
The final integral on the right in (10) must be individually adapted to the slab {\em{per se}} and to each of the
enclosing half spaces, its respective values being denoted as $I_{1}^{\pm}(z)$ for $z>0,$ $I_{2}^{\pm}(z)$ when $0>z>-a,$ and
finally $I_{3}^{\pm}(z)$ throughout the lower half space with $-a>z.$  We get:

\parindent=0in
{\bf{\underline{{\mbox{\boldmath$I_{1}^{\pm}(z),\; z\!>\!0$}}}:}}         
\parindent=0.5in
\begin{eqnarray}
I_{1}^{\pm}(z) & = & \exp\left(\rule{0mm}{3.5mm}\!-ik_{1}{\hat{n}}_{z}z\right)\int_{-a}^{\,0}
      \exp\left(\!\rule{0mm}{4mm}\left\{\rule{0mm}{3.5mm}ik_{1}{\hat{n}}_{z}+p_{\pm}\right\}\!z'\right)dz'  \nonumber \\
  & = & \exp\left(\rule{0mm}{3.5mm}\!-ik_{1}{\hat{n}}_{z}z\right)\left\{\rule{0mm}{3.5mm}ik_{1}{\hat{n}}_{z}+p_{\pm}\right\}^{-1}
  \left[\rule{0mm}{5mm}\,1-
  \exp\left(\!\rule{0mm}{4mm}-\left\{\rule{0mm}{3.5mm}ik_{1}{\hat{n}}_{z}+p_{\pm}\right\}\!a\right)\,\right] \,;    
\end{eqnarray}     

\parindent=0in
{\bf{\underline{{\mbox{\boldmath$I_{3}^{\pm}(z),\; -a\!>\!z$}}}:}}         
\parindent=0.5in
\begin{eqnarray}
I_{3}^{\pm}(z) & = &\rule{3mm}{0mm} \exp\left(\rule{0mm}{3.5mm}ik_{1}{\hat{n}}_{z}z\right)\int_{-a}^{\,0}
\exp\left(\!\rule{0mm}{4mm}-\left\{\rule{0mm}{3.5mm}ik_{1}{\hat{n}}_{z}-p_{\pm}\right\}\!z'\right)dz'  \nonumber \\
& = & -\exp\left(\rule{0mm}{3.5mm}ik_{1}{\hat{n}}_{z}z\right)\left\{\rule{0mm}{3.5mm}ik_{1}{\hat{n}}_{z}-p_{\pm}\right\}^{-1}
\left[\rule{0mm}{5mm}\,1-
\exp\left(\!\rule{0mm}{4mm}\left\{\rule{0mm}{3.5mm}ik_{1}{\hat{n}}_{z}-p_{\pm}\right\}\!a\right)\,\right] \,;    
\end{eqnarray}
and then

\parindent=0in
{\bf{\underline{{\mbox{\boldmath$I_{2}^{\pm}(z),\; 0\!>\!z\!>\!-a$}}}:}}         
\parindent=0.5in
\begin{eqnarray}
I_{2}^{\pm}(z) & = & \exp\left(\rule{0mm}{3.5mm}\!-ik_{1}{\hat{n}}_{z}z\right)\int_{-a}^{\,z}
\exp\left(\!\rule{0mm}{4mm}\left\{\rule{0mm}{3.5mm}ik_{1}{\hat{n}}_{z}+p_{\pm}\right\}\!z'\right)dz'  \nonumber \\
 &  & \rule{2cm}{0mm} + \exp\left(\rule{0mm}{3.5mm}ik_{1}{\hat{n}}_{z}z\right)\int_{\,z}^{\,0}
\exp\left(\!\rule{0mm}{4mm}-\left\{\rule{0mm}{3.5mm}ik_{1}{\hat{n}}_{z}-p_{\pm}\right\}\!z'\right)dz'   \\
 & = & \exp\left(\rule{0mm}{3.5mm}\!-ik_{1}{\hat{n}}_{z}z\right)\left\{\rule{0mm}{3.5mm}ik_{1}{\hat{n}}_{z}+p_{\pm}\right\}^{-1}
 \left[\rule{0mm}{5mm}\,\exp\left(\!\rule{0mm}{4mm}\left\{\rule{0mm}{3.5mm}ik_{1}{\hat{n}}_{z}+p_{\pm}\right\}\!z\right)-
 \exp\left(\!\rule{0mm}{4mm}-\left\{\rule{0mm}{3.5mm}ik_{1}{\hat{n}}_{z}+p_{\pm}\right\}\!a\right)\,\right] \nonumber \\
 &  & \rule{2cm}{0mm} 
    +\exp\left(\rule{0mm}{3.5mm}ik_{1}{\hat{n}}_{z}z\right)\left\{\rule{0mm}{3.5mm}ik_{1}{\hat{n}}_{z}-p_{\pm}\right\}^{-1}
 \left[\rule{0mm}{5mm}\,
 \exp\left(\!\rule{0mm}{4mm}-\left\{\rule{0mm}{3.5mm}ik_{1}{\hat{n}}_{z}-p_{\pm}\right\}\!z\right)-1\,\right] \,. \nonumber
\end{eqnarray}
\newpage
\mbox{   }
\newline

     On collating (11) and (12) with (10) we see that $I_{3}^{\pm}(z)$ automatically assembles a phase progression
$\exp\left(\rule{0mm}{3.5mm}{ik_{1}{\rm{\bf{\hat{n}}}}{\mbox{\boldmath$\cdot$}}{\bf{r}}}\right)$ befitting wave
penetration into the subjacent half space.  Structure $I_{1}^{\pm}(z)$ similarly provides for the emission of a
reflected wave governed by
$\exp\left(\rule{0mm}{3.25mm}ik_{1}\left\{{\rm{\bf{\hat{n}}}}_{\|}
        {\mbox{\boldmath$\cdot$}}{\mbox{\boldmath$\rho$}}-{\hat{n}}_{z}z\right\}\right)$ as its phase factor, a
traditional outcome which we formalize by setting
${\rm{\bf{\hat{n}}'}}={\rm{\bf{\hat{n}}}}${\scriptsize{$_{\|}$}}$-\,{\hat{n}}_{z}{\rm{\bf{\hat{e}}}}_{z}$ so as to
arrive finally at its abbreviated form
$\exp\left(\rule{0mm}{3.5mm}{ik_{1}{\rm{\bf{\hat{n}}'}}{\mbox{\boldmath$\cdot$}}{\bf{r}}}\right)\!.$  In both cases
the indicated wave progression is of a pure unidirectional type.

      By contrast, intermediate structure $I_{2}^{\pm}(z)$ from (13) mixes the inherent slab propagation according to
$\exp\left(\rule{0mm}{1.5mm}ik_{1}{\rm{\bf{\hat{n}}}}_{\|}{\mbox{\boldmath$\cdot$}}{\mbox{\boldmath$\rho$}}+ p_{\pm}z\right)$ 
with both exterior progression types found under (11) and (12).  We make this segregation into propagation types more explicit by rewriting (13) as

\parindent=0in
{\bf{\underline{{\mbox{\boldmath$I_{2}^{\pm}(z),\; 0\!>\!z\!>\!-a$}}}:}}         
\parindent=0.5in
\begin{eqnarray}
I_{2}^{\pm}(z)
 & = & \left\{\frac{1}{ik_{1}{\hat{n}}_{z}+p_{\pm}}+\frac{1}{ik_{1}{\hat{n}}_{z}-p_{\pm}}\right\}e^{p_{\pm}z} 
   -\frac{1}{ik_{1}{\hat{n}}_{z}-p_{\pm}}\,e^{ik_{1}{\hat{n}}_{z}z}
   -\frac{e^{-\{ik_{1}{\hat{n}}_{z}+p_{\pm}\}a}}{ik_{1}{\hat{n}}_{z}+p_{\pm}}\,e^{-ik_{1}{\hat{n}}_{z}z}
\end{eqnarray}             
wherein, for imminent use in (19), we note that
\begin{equation}  	
\frac{1}{ik_{1}{\hat{n}}_{z}+p_{\pm}}+\frac{1}{ik_{1}{\hat{n}}_{z}-p_{\pm}} =  
   -\frac{2ik_{1}{\hat{n}}_{z}}{k_{1}^{2}\,{\hat{n}}_{z}^{2}+p_{\pm}^{2}} =
    \frac{2k_{1}{\hat{n}}_{z}}{\omega\mu\left(\sigma_{2}-i\omega\{\epsilon_{2}-\epsilon_{1}\}\right)}\,,
\end{equation}
the final member on the right emerging after appeal to Eq. (2).  Imminent too in (20) is a need for the two statements that
\begin{equation}
{\bf{   {P}}^{\pm}}{\mbox{\boldmath$\cdot\,$}}{\bf{   {P}}^{\pm}}=p_{\pm}^{2}
-k_{1}^{2}{\rm{\bf{\hat{n}}}}_{\|}{\mbox{\boldmath$\cdot\,$}}{\rm{\bf{\hat{n}}}}_{\|}=
-\omega^{2}\mu\epsilon_{2}\left(\rule{0mm}{4mm}1+i\sigma_{2}/\omega\epsilon_{2}\right)\,,
\end{equation}
statements already anticipated by (2).
\section{Up/down half space field structure}
     With the tools (10)-(12) in hand, we can at once state the total half space fields respectively above and below
in the highly symmetric forms
{\small{
\begin{eqnarray}
\rule{-5mm}{0mm}{\bf{  {E}}}_{{\rm{tot}}}({\bf{r}}) &\!\!\! = \!\!\! &
	{\bf{  {E}}}_{{\rm{inc}}}e^{ik_{1}{\rm{\bf{\hat{n}}}}{\mbox{\boldmath$\cdot$}}{\bf{r}}}  
 -\frac{k_{1}(\sigma_{2}-i\omega\{\epsilon_{2}-\epsilon_{1}\})}{2{\hat{n}}_{z}\epsilon_{1}\omega}\,
e^{ik_{1}{\rm{\bf{\hat{n}}'}}{\mbox{\boldmath$\cdot$}}{\bf{r}}}\,
{\rm{\bf{\hat{n}}'}}{\mbox{\boldmath$\,\times$}}
\left[\rule{0mm}{5mm}\,{\rm{\bf{\hat{n}}'}}{\mbox{\boldmath$\,\times\,$}}\sum_{\pm}\left\{\frac{1-
	\exp\left(\!\rule{0mm}{4mm}-\left\{\rule{0mm}{3.5mm}ik_{1}{\hat{n}}_{z}+p_{\pm}\right\}\!a\right)}
{ik_{1}{\hat{n}}_{z}+p_{\pm}}\right\}{\bf{  {E}}}_{{\rm{slab}}}^{\pm}\,\right] 
\end{eqnarray} }}
and
{\small{
\begin{eqnarray}
\rule{3mm}{0mm}
{\bf{  {E}}}_{{\rm{tot}}}({\bf{r}}) &\!\!\!=\!\!\! & e^{ik_{1}{\rm{\bf{\hat{n}}}}{\mbox{\boldmath$\cdot$}}{\bf{r}}}
\left(\rule{0mm}{8mm}{\bf{  {E}}}_{{\rm{inc}}}
+\frac{k_{1}(\sigma_{2}-i\omega\{\epsilon_{2}-\epsilon_{1}\})}{2{\hat{n}}_{z}\epsilon_{1}\omega}\,
{\rm{\bf{\hat{n}}}}{\mbox{\boldmath$\,\times$}}\!
\left[\rule{0mm}{5mm}\,{\rm{\bf{\hat{n}}}}{\mbox{\boldmath$\,\times\,$}}\!\sum_{\pm}\left\{\frac{1-
	\exp\left(\!\rule{0mm}{4mm}\left\{\rule{0mm}{3.5mm}ik_{1}{\hat{n}}_{z}-p_{\pm}\right\}\!a\right)}
{ik_{1}{\hat{n}}_{z}-p_{\pm}}\right\}{\bf{  {E}}}_{{\rm{slab}}}^{\pm}\,\right] \right).
\end{eqnarray} }}
\newpage
\mbox{   }
\newline
\newline
\newline
\parindent=0in
Not to be overlooked is the automatic transversality to propagation directions ${\rm{\bf{\hat{n}}'}}$ and
${\rm{\bf{\hat{n}}}}$ with which (17) and (18) endow the up/down radiated field contributions.  At the same time
the information which (17) and (18) provide remains incomplete until such time as vector amplitudes 
${\bf{  {E}}}_{{\rm{slab}}}^{\pm}$ have been fully ascertained.  Sections 7 and 8, Eqs. (30), (31), (33), and
(37) below provide the requisite completion.
\parindent=0.5in
\vspace{-4mm}
\section{Solving the slab integral equation:  step 1}
\vspace{-3mm} 
     With reference to (1), (5), (10), (14)-(15), Eq. (9), when stated within the slab interior $0>z>-a,$
can now be written as
\begin{eqnarray}
\lefteqn{
\left(\frac{\epsilon_{2}+i\sigma_{2}/\omega}{\epsilon_{1}}\right)\sum_{\pm}{\bf{  {E}}}_{{\rm{slab}}}^{\pm}
\exp\left(\rule{0mm}{3.5mm}{\bf{   {P}}^{\pm}}{\mbox{\boldmath$\cdot\,$}}{\rm{\bf{r}}}\right)
  =  {\bf{  {E}}}_{{\rm{inc}}}e^{ik_{1}{\rm{\bf{\hat{n}}}}{\mbox{\boldmath$\cdot$}}{\bf{r}}}  }\nonumber  \\
 &   &\rule{1.5cm}{0mm} + k_{1}^{-2}\sum_{\pm}{\bf{   {P}}^{\pm}}{\mbox{\boldmath$\times$}}
\left\{\rule{0mm}{5mm}{\bf{   {P}}^{\pm}}{\mbox{\boldmath$\times\,$}}{\bf{  {E}}}_{{\rm{slab}}}^{\pm}\right\}
\exp\left(\rule{0mm}{3.5mm}{\bf{   {P}}^{\pm}}{\mbox{\boldmath$\cdot\,$}}{\rm{\bf{r}}}\right) \nonumber  \\
 &  &\rule{1.5cm}{0mm}+\,\left(\frac{k_{1}\left(\sigma_{2}-i\omega\left\{\epsilon_{2}-\epsilon_{1}\right\}\right)} {2{\hat{n}}_{z}\epsilon_{1}\omega}\right)\,
  e^{ik_{1}{\rm{\bf{\hat{n}}}}{\mbox{\boldmath$\cdot$}}{\bf{r}}}\,\,
{\rm{\bf{\hat{n}}}}{\mbox{\boldmath$\,\times$}}\!
\left[\rule{0mm}{5mm}\,{\rm{\bf{\hat{n}}}}{\mbox{\boldmath$\,\times\,$}}\!\sum_{\pm}\left\{\frac{1}	
{ik_{1}{\hat{n}}_{z}-p_{\pm}}\right\}{\bf{  {E}}}_{{\rm{slab}}}^{\pm}\,\right]   \\
 &   &\rule{1.5cm}{0mm} +\left(\frac{k_{1}\left(\sigma_{2}-i\omega\left\{\epsilon_{2}-\epsilon_{1}\right\}\right)} {2{\hat{n}}_{z}\epsilon_{1}\omega}\right) 
e^{ik_{1}{\rm{\bf{\hat{n}'}}}{\mbox{\boldmath$\cdot\,$}}{\bf{r}}}\,
{\rm{\bf{\hat{n}'}}}{\mbox{\boldmath$\,\times$}}\!
\left[\rule{0mm}{5mm}\,{\rm{\bf{\hat{n}'}}}{\mbox{\boldmath$\,\times\,$}}\!\sum_{\pm}
\left\{\frac{\exp\left(\!\rule{0mm}{4mm}-\left\{\rule{0mm}{3.5mm}ik_{1}{\hat{n}}_{z}+p_{\pm}\right\}\!a\right)}	
{ik_{1}{\hat{n}}_{z}+p_{\pm}}\right\}{\bf{  {E}}}_{{\rm{slab}}}^{\pm}\,\right].  \nonumber
\end{eqnarray}
And then, since
\begin{eqnarray}
{\bf{   {P}}^{\pm}}{\mbox{\boldmath$\times$}}
\left\{\rule{0mm}{5mm}{\bf{   {P}}^{\pm}}{\mbox{\boldmath$\times\,$}}{\bf{  {E}}}_{{\rm{slab}}}^{\pm}\right\} & = &
\left({\bf{   {P}}^{\pm}}{\mbox{\boldmath$\cdot\,$}}{\bf{  {E}}}_{{\rm{slab}}}^{\pm}\right){\bf{   {P}}^{\pm}}-
\left({\bf{   {P}}^{\pm}}{\mbox{\boldmath$\cdot\,$}}{\bf{   {P}}}^{\pm}\right)
            {\bf{  {E}}_{{\rm{slab}}}^{\pm}}\nonumber \\
   & = & k_{1}^{2}\left(\frac{\epsilon_{2}+i\sigma_{2}/\omega}{\epsilon_{1}}\right) {\bf{  {E}}_{{\rm{slab}}}^{\pm}}\,,        
\end{eqnarray}
the second line holding in consequence of Eqs. (4) and (16), we see that the slab field
$\sum_{\pm}{\bf{  {E}}}_{{\rm{slab}}}^{\pm}
\exp\left(\rule{0mm}{3.5mm}{\bf{   {P}}^{\pm}}{\mbox{\boldmath$\cdot\,$}}{\rm{\bf{r}}}\right)$ {\em{per se}}
cancels\footnote{Cancellation of this sort is a persistent analytic phenomenon uniformly encountered
whenever analogues of integral equation (9) are brought to bear on other geometries.  In particular, it
occurs in the case of a dissipative sphere exposed to a punctual electric dipole radiating in close
proximity.  Solution details, albeit inherently intricate, are, from a conceptual viewpoint, no more
involved than those here undertaken.  A presentation, complete with a Poynting/Joule energy budget
analysis, is set out in two unpublished internal memoranda, proprietary to Allwave Corporation and dated on April 29
and May 19, 2009.

      It would be tempting to muster out similar energy assessments in the present planar context.  But we shall
refrain from doing so, first in the interest of maintaining some semblance of brevity, and second, since,
in fact, such assessments, at base, have nothing whatsoever to do with the electromagnetic thesis at hand,
a thesis that centers on radiative self-consistency.      

      Moreover, the fact that integration over slab polarization/ohmic sources not only reproduces the internal
field exactly, without in any way subjecting it directly to additional conditions, a field further accompanied by
type 1 contributions which latter must then neutralize the incoming, type 1 excitation, as conveyed by Eqs. (22)-(23),
can be regarded as a manifestation of the Ewald-Oseen extinction theorem, even though the present analysis differs
radically from the proof on view in [{\bf{4}}].}
identically from both sides of (19), leaving us to contend with just
\begin{eqnarray}
{\bf{0}} & = & \rule{5mm}{0mm}
 {\bf{  {E}}}_{{\rm{inc}}}e^{ik_{1}{\rm{\bf{\hat{n}}}}{\mbox{\boldmath$\cdot$}}{\bf{r}}}  \nonumber  \\
 &   &\rule{0cm}{0mm} + \,\left(\frac{k_{1}\left(\sigma_{2}-i\omega\left\{\epsilon_{2}-\epsilon_{1}\right\}\right)} {2{\hat{n}}_{z}\epsilon_{1}\omega}\right)\,
e^{ik_{1}{\rm{\bf{\hat{n}}}}{\mbox{\boldmath$\cdot$}}{\bf{r}}}\,\,
{\rm{\bf{\hat{n}}}}{\mbox{\boldmath$\,\times$}}\!
\left[\rule{0mm}{5mm}\,{\rm{\bf{\hat{n}}}}{\mbox{\boldmath$\,\times\,$}}\!\sum_{\pm}\left\{\frac{1}	
{ik_{1}{\hat{n}}_{z}-p_{\pm}}\right\}{\bf{  {E}}}_{{\rm{slab}}}^{\pm}\,\right]   \\
 &   &\rule{0cm}{0mm} + \left(\frac{k_{1}\left(\sigma_{2}-i\omega\left\{\epsilon_{2}-\epsilon_{1}\right\}\right)} {2{\hat{n}}_{z}\epsilon_{1}\omega}\right) 
e^{ik_{1}{\rm{\bf{\hat{n}'}}}{\mbox{\boldmath$\cdot\,$}}{\bf{r}}}\,
{\rm{\bf{\hat{n}'}}}{\mbox{\boldmath$\,\times$}}\!
\left[\rule{0mm}{5mm}\,{\rm{\bf{\hat{n}'}}}{\mbox{\boldmath$\,\times\,$}}\!\sum_{\pm}
\left\{\frac{\exp\left(\!\rule{0mm}{4mm}-\left\{\rule{0mm}{3.5mm}ik_{1}{\hat{n}}_{z}+p_{\pm}\right\}\!a\right)}	
{ik_{1}{\hat{n}}_{z}+p_{\pm}}\right\}{\bf{  {E}}}_{{\rm{slab}}}^{\pm}\,\right].  \nonumber
\end{eqnarray}
But, as the functions $e^{\pm ik_{1}{\hat{n}}_{z}z}$ are linearly independent, it follows that (21)
decouples into the two individual vector statements
\begin{eqnarray} 
{\rm{\bf{\hat{n}}}}{\mbox{\boldmath$\,\times$}}\!
\left[\rule{0mm}{5mm}\,{\rm{\bf{\hat{n}}}}{\mbox{\boldmath$\,\times\,$}}\!\sum_{\pm}\left\{\frac{1}	
{ik_{1}{\hat{n}}_{z}-p_{\pm}}\right\}{\bf{  {E}}}_{{\rm{slab}}}^{\pm}\,\right] & = &
  -\left(\frac{2{\hat{n}}_{z}\epsilon_{1}\omega}
  {k_{1}\left(\sigma_{2}-i\omega\left\{\epsilon_{2}-\epsilon_{1}\right\}\right)}\right){\bf{  {E}}}_{{\rm{inc}}} 
\end{eqnarray}
and
\begin{eqnarray}  
{\rm{\bf{\hat{n}'}}}{\mbox{\boldmath$\,\times$}}\!
\left[\rule{0mm}{5mm}\,{\rm{\bf{\hat{n}'}}}{\mbox{\boldmath$\,\times\,$}}\!\sum_{\pm}
\left\{\frac{\exp\left(\!\rule{0mm}{4mm}-\left\{\rule{0mm}{3.5mm}ik_{1}{\hat{n}}_{z}+p_{\pm}\right\}\!a\right)}	
{ik_{1}{\hat{n}}_{z}+p_{\pm}}\right\}{\bf{  {E}}}_{{\rm{slab}}}^{\pm}\,\right] & = & {\bf{0}}\,,
\end{eqnarray}
\newpage
\mbox{    }
\newline  
\newline
\newline
fully equivalent to six equations for the six components of vectors ${\bf{  {E}}}_{{\rm{slab}}}^{\pm}.$
\section{Solving the slab integral equation:  step 2}
     We tackle first the homogeneous, subsidiary condition (23).  But before doing so we ease and systematize
somewhat the ensuing notation by setting
\begin{equation}
\left\{\rule{0mm}{2.0cm}\begin{array}{rcl}
    {\bf{   {P}}^{\,\nu}}{\mbox{\boldmath$\cdot\,$}}{\rm{\bf{\hat{n}}}} \rule{0.8mm}{0mm} & = & \rule{3mm}{0mm}
                       ik_{1}\sin^{2}\vartheta - p_{\nu}\cos\vartheta   \\
                             &     &     \\
       {\bf{   {P}}^{\,\nu}}{\mbox{\boldmath$\cdot\,$}}{\rm{\bf{\hat{n}'}}} & = &  \rule{3mm}{0mm}
                       ik_{1}\sin^{2}\vartheta + p_{\nu}\cos\vartheta   \\
                             &     &     \\
       \gamma_{\nu} & = & -ik_{1}\cos\vartheta+p_{\nu}    \\  
                             &     &     \\
               \nu & = & \rule{3mm}{0mm} \pm \,.            
        \end{array} \right.                                     
\end{equation}
In particular, on collating the last two denominators from (15), we observe at once that
\begin{equation}
\omega\mu\left(\sigma_{2}-i\omega\left\{\epsilon_{2}-\epsilon_{1}\right\}\right)=
i\left(k_{1}^{2}\,{\hat{n}}_{z}^{2}+p_{\pm}^{2}\right)=i\left(k_{1}^{2}\cos^{2}\vartheta+p_{\pm}^{2}\right)=
-i\gamma_{\mp}\gamma_{\pm}\,.
\end{equation}
And, suffer though it may from the self-evident   
redundancies ${\bf{   {P}}^{-\nu}}{\mbox{\boldmath$\cdot\,$}}{\rm{\bf{\hat{n}}}}=            	
{\bf{   {P}}^{\,\nu}}{\mbox{\boldmath$\cdot\,$}}{\rm{\bf{\hat{n}'}}},$
tableau (24) will nevertheless be immediately seen to bestow a much valued notational compression.
In any event, it now follows from (23), once its repeated vector product has been duly unraveled, that
\begin{eqnarray}
\sum_{\nu}\gamma_{\nu}^{-1}e^{-\,\gamma_{\nu}a}
        {\bf{  {E}}}_{{\rm{slab}}}^{\,\nu}& = & \alpha\,{\rm{\bf{\hat{n}'}}}\,,
\end{eqnarray}
with coefficient $\alpha$ yet to be determined.  And, with a view to (4), we then further get
\begin{equation}
{\bf{   {P}}^{\,\nu}}{\mbox{\boldmath$\cdot\,$}}{\bf{  {E}}}_{{\rm{slab}}}^{-\nu}= \alpha\,
 \gamma_{-\nu}e^{\gamma_{-\nu}a}\left(\rule{0mm}{4mm}{\bf{   {P}}^{\,\nu}}
    {\mbox{\boldmath$\cdot\,$}}{\rm{\bf{\hat{n}'}}}\right)\,.
\end{equation}

     In similar fashion, scalar multiplication by ${\bf{   {P}}^{\,\nu}}$ of the inhomogeneous relation (22),
taking advantage once again of (4), yields
\begin{eqnarray}
\sum_{\nu'}\gamma_{-\nu'}^{-1}\left({\rm{\bf{\hat{n}}}}{\mbox{\boldmath$\,\cdot\,$}}
{\bf{  {E}}}_{{\rm{slab}}}^{\nu'}\right) & = &   \nonumber \\
 &   & \rule{-3cm}{0mm} = \left(\rule{0mm}{3.5mm}{\bf{   {P}}^{\,\nu}}{\mbox{\boldmath$\cdot\,$}}{\rm{\bf{\hat{n}}}}\right)^{\!-1}\left\{\rule{0mm}{7mm}
 \gamma_{\nu}^{-1}{\bf{   {P}}^{\,\nu}}{\mbox{\boldmath$\cdot\,$}}{\bf{  {E}}}_{{\rm{slab}}}^{-\nu}
 -\left(\frac{2{\hat{n}}_{z}\epsilon_{1}\omega}
 {k_{1}\left(\sigma_{2}-i\omega\left\{\epsilon_{2}-\epsilon_{1}\right\}\right)}\right)
 {\bf{   {P}}^{\,\nu}}{\mbox{\boldmath$\cdot\,$}}{\bf{  {E}}}_{{\rm{inc}}}\right\}\,.  
\end{eqnarray}
We subtract next the two versions of (28) at $\nu=\pm$ so as to remove its left member, and we bring in the information
already acquired in (27), producing thereby an explicit, if seemingly unwieldy determination of coefficient $\alpha.$
Thus
\begin{eqnarray}
\alpha\sum_{\nu}\nu\,\frac{\gamma_{-\nu}e^{\gamma_{-\nu}a}\left(\rule{0mm}{4mm}{\bf{   {P}}^{\,\nu}}
	{\mbox{\boldmath$\cdot\,$}}{\rm{\bf{\hat{n}'}}}\right)}
{\gamma_{\nu}\left(\rule{0mm}{4mm}{\bf{   {P}}^{\,\nu}}{\mbox{\boldmath$\cdot\,$}}{\rm{\bf{\hat{n}}}}\right)}
  & = & \left(\frac{2{\hat{n}}_{z}\epsilon_{1}\omega}
  {k_{1}\left(\sigma_{2}-i\omega\left\{\epsilon_{2}-\epsilon_{1}\right\}\right)}\right)
  \sum_{\nu}\nu\,\frac{\left(\rule{0mm}{4mm}{\bf{   {P}}^{\,\nu}}{\mbox{\boldmath$\cdot\,$}}
  	{\bf{  {E}}}_{{\rm{inc}}}\right)}
  {\left(\rule{0mm}{4mm}{\bf{   {P}}^{\,\nu}}{\mbox{\boldmath$\cdot\,$}}{\rm{\bf{\hat{n}}}}\right)}\,.  
\end{eqnarray}
This determination emerges finally in the more streamlined form
\newpage
\mbox{   }
\newline
\begin{eqnarray}
\alpha\sum_{\nu}\nu\,\frac{\gamma_{-\nu}^{2}e^{\gamma_{-\nu}a}\left(\rule{0mm}{4mm}{\bf{   {P}}^{\,\nu}}
	{\mbox{\boldmath$\cdot\,$}}{\rm{\bf{\hat{n}'}}}\right)}
{\left(\rule{0mm}{4mm}{\bf{   {P}}^{\,\nu}}{\mbox{\boldmath$\cdot\,$}}{\rm{\bf{\hat{n}}}}\right)}
& = & 2ik_{1}{\hat{n}}_{z}
\sum_{\nu}\nu\,\frac{\left(\rule{0mm}{4mm}{\bf{   {P}}^{\,\nu}}{\mbox{\boldmath$\cdot\,$}}
	{\bf{  {E}}}_{{\rm{inc}}}\right)}
{\left(\rule{0mm}{4mm}{\bf{   {P}}^{\,\nu}}{\mbox{\boldmath$\cdot\,$}}{\rm{\bf{\hat{n}}}}\right)}\,.  
\end{eqnarray}
when further use is made of (25).
\section{Solving the slab integral equation:  step 3}
      Since $\alpha$ is now known, we can navigate freely between (22) and (26) so as to finalize an explicit
buildup of both slab amplitudes ${\bf{  {E}}}_{{\rm{slab}}}^{\,\pm}.$  For instance, setting, on the basis
of (26),
\begin{equation}
{\bf{  {E}}}_{{\rm{slab}}}^{\,-} = \gamma_{-}e^{\gamma_{-}a}\left\{\rule{0mm}{6mm}
\alpha {\rm{\bf{\hat{n}'}}}-\gamma_{+}^{-1}e^{-\,\gamma_{+}a}{\bf{  {E}}}_{{\rm{slab}}}^{\,+}\right\}\,,
\end{equation}
its subsequent introduction into (22) yields
\begin{eqnarray} 
{\rm{\bf{\hat{n}}}}{\mbox{\boldmath$\,\times$}}\!
\left[\rule{0mm}{6mm}\,{\rm{\bf{\hat{n}}}}{\mbox{\boldmath$\,\times\,$}}\!\left(\rule{0mm}{7mm}	
\alpha\gamma_{+}\gamma_{-}^{2}e^{\gamma_{-}a}{\rm{\bf{\hat{n}'}}}+\left\{\rule{0mm}{6mm}
\gamma_{+}^{2}-\gamma_{-}^{2}e^{(\gamma_{-}-\,\gamma_{+})a}\right\}
{\bf{  {E}}}_{{\rm{slab}}}^{+}\,\right)\,\right] & = &
-2ik_{1}{\hat{n}}_{z}\gamma_{+}{\bf{  {E}}}_{{\rm{inc}}} \,.
\end{eqnarray}
And then, on segregating the several vectorial terms of (32) into those already known
\begin{eqnarray}
{\bf{\Omega}} & = & i\left\{\rule{0mm}{6mm}\gamma_{-}^{2}e^{(\gamma_{-}-\,\gamma_{+})a}-\,\gamma_{+}^{2}\right\}^{\!-1}\!
    \left[\rule{0mm}{7mm}\,2k_{1}{\hat{n}}_{z}\gamma_{+}{\bf{  {E}}}_{{\rm{inc}}}-
    i\alpha\gamma_{+}\gamma_{-}^{2}e^{\gamma_{-}a}{\rm{\bf{\hat{n}}}}{\mbox{\boldmath$\,\times$}}\!
    \left(\rule{0mm}{4.5mm}\,{\rm{\bf{\hat{n}}}}{\mbox{\boldmath$\,\times\,$}}{\rm{\bf{\hat{n}'}}}\right)\right]\,,
\end{eqnarray}
and those {\em{a priori}} presumed to be still unknown, {\em{viz.,}} ${\rm{\bf{\hat{n}}}}{\mbox{\boldmath$\,\times$}}\!
\left(\rule{0mm}{4.5mm}\,{\rm{\bf{\hat{n}}}}{\mbox{\boldmath$\,\times\,$}}{\bf{  {E}}}_{{\rm{slab}}}^{\,+}\right)\!,$
we nonetheless find
\begin{equation}
{\rm{\bf{\hat{n}}}}{\mbox{\boldmath$\,\times$}}\!
\left(\rule{0mm}{4.5mm}\,{\rm{\bf{\hat{n}}}}{\mbox{\boldmath$\,\times\,$}}{\bf{  {E}}}_{{\rm{slab}}}^{\,+}\right)={\bf{\Omega}}\,,
\end{equation}
or else
\begin{equation}
{\rm{\bf{\hat{n}}}}\left(\rule{0mm}{4mm}{\rm{\bf{\hat{n}}}}{\mbox{\boldmath$\,\cdot\,$}}{\bf{  {E}}}_{{\rm{slab}}}^{+}\right)-
{\bf{  {E}}}_{{\rm{slab}}}^{+} = {\bf{\Omega}}\,,
\end{equation}
whereupon an appeal yet again to (4) now yields
\begin{equation}
\left(\rule{0mm}{4mm}{\bf{   {P}}^{+}}{\mbox{\boldmath$\!\cdot\,$}}{\rm{\bf{\hat{n}}}}\right)
\left(\rule{0mm}{4mm}{\rm{\bf{\hat{n}}}}{\mbox{\boldmath$\,\cdot\,$}}{\bf{  {E}}}_{{\rm{slab}}}^{+}\right)=
\left(\rule{0mm}{4mm}{\bf{   {P}}^{+}}{\mbox{\boldmath$\!\cdot\,$}}{\bf{\Omega}}\right)
\end{equation}
and thus fixes
\begin{equation}
{\bf{  {E}}}_{{\rm{slab}}}^{+}={\rm{\bf{\hat{n}}}}\,
\frac{\left(\rule{0mm}{4mm}{\bf{   {P}}^{+}}{\mbox{\boldmath$\!\cdot\,$}}{\bf{\Omega}}\right)}
{\left(\rule{0mm}{4mm}{\bf{   {P}}^{+}}{\mbox{\boldmath$\!\cdot\,$}}{\rm{\bf{\hat{n}}}}\right)}-{\bf{\Omega}}
\end{equation}
in its entirety.  A retrospective glance at (31) confirms that full knowledge regarding ${\bf{  {E}}}_{{\rm{slab}}}^{+}$
fixes likewise partner ${\bf{  {E}}}_{{\rm{slab}}}^{-}$ and, in principle, we are done.
\newpage
\mbox{   }
\section{A partial reconciliation with boundary matching}
      As initially announced, we have now succeeded in solving the entire slab reflection/transmission problem without
ever once allowing our pen to stray in the direction of boundary matching and its determinantal support apparatus.  Still, it
behooves us to demonstrate some level of concordance, however minimal.

      We propose indeed to base a technique comparison in the very simplest context, that of normal wave incidence.
While admittedly this will not exercise the full scope of either apparatus, avoidance of a potential failure
should weaken any {\em{a priori}} urge to condemn.  But, as will happily turn out, our limited comparison is destined
to emerge unscathed by any such failure.

       At normal incidence $\vartheta=0,$ ${\rm{\bf{\hat{n}}}}=-{\rm{\bf{\hat{e}}}}_{z},$ ${\hat{n}}_{z}=-1,$ so that
${\bf{   {P}}^{\pm}}{\mbox{\boldmath$\cdot\,$}}{\bf{  {E}}}_{{\rm{inc}}}=0,$ whence $\alpha=0$ and thus
\begin{equation}
{\bf{  {E}}}_{{\rm{slab}}}^{\,-} =
 -\gamma_{-}\gamma_{+}^{-1}e^{\left(\gamma_{-}-\,\gamma_{+}\right)a}{\bf{  {E}}}_{{\rm{slab}}}^{\,+}\,.
\end{equation}
As a replacement of (3) we now have
\begin{eqnarray}
p_{\pm} & = & \pm\frac{k_{1}}{\sqrt{\rule{0mm}{3mm}2\epsilon_{1}\,}} \left\{\rule{0mm}{8mm}
\sqrt{\sqrt{\rule{0mm}{4mm}\epsilon_{2}^{2}+
		\left(\rule{0mm}{4mm}\sigma_{2}/\omega\right)^{2}\,}-\epsilon_{2}\,}\,  
 -i\frac{\omega}{|\,\omega\,|}
\sqrt{\sqrt{\rule{0mm}{4mm}\epsilon_{2}^{2}+
		\left(\rule{0mm}{4mm}\sigma_{2}/\omega\right)^{2}\,}+\epsilon_{2}\,}  \;\right\}			
\end{eqnarray}
and $\gamma_{\pm}=-ik_{1}+p_{\pm}.$ Moreover
${\rm{\bf{\hat{n}}}}{\mbox{\boldmath$\,\times\,$}}{\rm{\bf{\hat{n}'}}}={\bf{0}}$ so that
\begin{eqnarray}
{\bf{\Omega}} & = & 2ik_{1}\gamma_{+}\left\{\rule{0mm}{6mm}\gamma_{+}^{2}-\gamma_{-}^{2}e^{(\gamma_{-}-\,\gamma_{+})a}\right\}^{\!-1}\!
{\bf{  {E}}}_{{\rm{inc}}}\,.
\end{eqnarray}
whereas
\begin{equation}
{\bf{  {E}}}_{{\rm{slab}}}^{+}={\rm{\bf{\hat{n}}}}\,
\frac{\left(\rule{0mm}{4mm}{\bf{   {P}}^{+}}{\mbox{\boldmath$\!\cdot\,$}}{\bf{\Omega}}\right)}
{\left(\rule{0mm}{4mm}{\bf{   {P}}^{+}}{\mbox{\boldmath$\!\cdot\,$}}{\rm{\bf{\hat{n}}}}\right)}-{\bf{\Omega}} =-{\bf{\Omega}}\,.
\end{equation}

       Turning first to the reflected vector amplitude ${\bf{  {E}}}_{{\rm{ref}}},$ we have from (17)
\begin{eqnarray}
{\bf{  {E}}}_{{\rm{ref}}} & = & -\frac{k_{1}(\sigma_{2}
-i\omega\{\epsilon_{2}-\epsilon_{1}\})}{2{\hat{n}}_{z}\epsilon_{1}\omega}\,\,
	{\rm{\bf{\hat{n}}'}}{\mbox{\boldmath$\,\times$}}
	\left[\rule{0mm}{5mm}\,{\rm{\bf{\hat{n}}'}}{\mbox{\boldmath$\,\times\,$}}\sum_{\pm}\left\{\frac{1-
		\exp\left(\!\rule{0mm}{4mm}-\left\{\rule{0mm}{3.5mm}ik_{1}{\hat{n}}_{z}+p_{\pm}\right\}\!a\right)}
	{ik_{1}{\hat{n}}_{z}+p_{\pm}}\right\}{\bf{  {E}}}_{{\rm{slab}}}^{\pm}\,\right] \nonumber   \\
	& = & -i\frac{\gamma_{-}\gamma_{+}}{2k_{1}}\,
	{\rm{\bf{\hat{n}}'}}{\mbox{\boldmath$\,\times$}}
	\left[\rule{0mm}{5mm}\,{\rm{\bf{\hat{n}}'}}{\mbox{\boldmath$\,\times\,$}}\sum_{\pm}\left\{\frac{1}		
	{ik_{1}{\hat{n}}_{z}+p_{\pm}}\right\}{\bf{  {E}}}_{{\rm{slab}}}^{\pm}\,\right]\,, 
\end{eqnarray}
the latter simplification gotten by virtue of (23).  But now, since clearly ${\rm{\bf{\hat{n}}'}}=-{\rm{\bf{\hat{n}}}}$
whereas (40), (41), and (38) insist that ${\bf{  {E}}}_{{\rm{slab}}}^{\pm}$ be both aligned along
${\bf{  {E}}}_{{\rm{inc}}},$ we further get
\begin{equation}
{\bf{  {E}}}_{{\rm{ref}}} = i\frac{\gamma_{-}\gamma_{+}}{2k_{1}}\,
\sum_{\pm}\left\{\frac{1}		
{ik_{1}{\hat{n}}_{z}+p_{\pm}}\right\}{\bf{  {E}}}_{{\rm{slab}}}^{\pm}=
\frac{i}{2k_{1}}\,\sum_{\pm}\gamma_{\mp}{\bf{  {E}}}_{{\rm{slab}}}^{\pm}
\end{equation}
which, written somewhat more explicitly, reads
\begin{eqnarray}
{\bf{  {E}}}_{{\rm{ref}}} & = & \gamma_{-}\gamma_{+}\left(\frac{1-e^{\left(\gamma_{-}-\,\gamma_{+}\right)a}}
{\gamma_{+}^{2}-\gamma_{-}^{2}e^{\left(\gamma_{-}-\,\gamma_{+}\right)a}}\right){\bf{  {E}}}_{{\rm{inc}}}\,.
\end{eqnarray}                              
\newpage
\mbox{   }
\newline

    From (18) and (22) combined we similarly find the total transmitted field vector amplitude
${\bf{  {E}}}_{{\rm{trans}}}$ in the form
\begin{eqnarray}
{\bf{  {E}}}_{{\rm{trans}}} & = &
{\bf{  {E}}}_{{\rm{inc}}}
+\frac{k_{1}(\sigma_{2}-i\omega\{\epsilon_{2}-\epsilon_{1}\})}{2{\hat{n}}_{z}\epsilon_{1}\omega}\,
{\rm{\bf{\hat{n}}}}{\mbox{\boldmath$\,\times$}}\!
\left[\rule{0mm}{5mm}\,{\rm{\bf{\hat{n}}}}{\mbox{\boldmath$\,\times\,$}}\!\sum_{\pm}\left\{\frac{1-
	\exp\left(\!\rule{0mm}{4mm}\left\{\rule{0mm}{3.5mm}ik_{1}{\hat{n}}_{z}-p_{\pm}\right\}\!a\right)}
{ik_{1}{\hat{n}}_{z}-p_{\pm}}\right\}{\bf{  {E}}}_{{\rm{slab}}}^{\pm}\,\right] \nonumber  \\
 & = & \frac{i}{2k_{1}}\sum_{\pm}\gamma_{\pm}e^{\gamma_{\mp}a}{\bf{  {E}}}_{{\rm{slab}}}^{\pm} \\
 & = & \left(\frac{\gamma_{+}^{2}-\gamma_{-}^{2}}
 {\gamma_{+}^{2}-\gamma_{-}^{2}e^{\left(\gamma_{-}-\,\gamma_{+}\right)a}}\right)
 e^{\gamma_{-}a}{\bf{  {E}}}_{{\rm{inc}}}\,.  \nonumber      
\end{eqnarray}
We pause to remark, by way of a partial check, that slab removal, signaled by having $a\downarrow 0+,$ implies
the requisite limits ${\bf{  {E}}}_{{\rm{ref}}}\rightarrow {\bf{0}}$ and
${\bf{  {E}}}_{{\rm{trans}}}\rightarrow {\bf{  {E}}}_{{\rm{inc}}}.$

      The standard treatment of this scattering problem insists upon interface continuity at $z=0$ and $z=-a$
of both electric and magnetic tangential field components, yielding altogether four linear equations for the four
vector amplitudes ${\bf{  {E}}}_{{\rm{ref}}},$ ${\bf{  {E}}}_{{\rm{trans}}},$ and ${\bf{  {E}}}_{{\rm{slab}}}^{\pm},$
all four of them coplanar and perpendicular to propagation vector ${\rm{\bf{\hat{n}}}}.$  Faraday's equation
selectively invoked according to vertical co{\"{o}}rdinate $z$ provides the magnetic field ${\bf{  {B}}}(z)$ in the
form
\begin{equation}
{\bf{  {B}}}(z)=\left\{\rule{0mm}{1.9cm}\begin{array}{lcl}
 -(k_{1}/\omega)\,{\rm{\bf{\hat{e}}}}_{z}{\mbox{\boldmath$\,\times\,$}}\!
 \left(\rule{0mm}{5mm}e^{-ik_{1}z}{\bf{  {E}}}_{{\rm{inc}}}-e^{ik_{1}z}{\bf{  {E}}}_{{\rm{ref}}}\right) & ; & z>0 \\
                        &      &      \\
 -i(p_{+}/\omega)\,{\rm{\bf{\hat{e}}}}_{z}{\mbox{\boldmath$\,\times\,$}}\!
 \left(\rule{0mm}{5mm}e^{p_{+}z}{\bf{  {E}}}_{{\rm{slab}}}^{+}-e^{-p_{+}z}{\bf{  {E}}}_{{\rm{slab}}}^{-}\right) & ; & 0>z>-a \\
                       &       &      \\
 -(k_{1}/\omega)\,{\rm{\bf{\hat{e}}}}_{z}{\mbox{\boldmath$\,\times\,$}}\!
 \left(\rule{0mm}{5mm}e^{-ik_{1}z}{\bf{  {E}}}_{{\rm{trans}}}\right) & ; & -a>z                       
                                 \end{array} \right. \,.
\end{equation}
Enforcement across the board of one additional cross product with vector ${\rm{\bf{\hat{e}}}}_{z}$ rotates
${\bf{  {B}}}({\bf{r}})$ into the common plane of electric vectors and supplies a vector quantity
${\rm{\bf{\hat{e}}}}_{z}{\mbox{\boldmath$\times\,$}}{\bf{  {B}}}(z)$ whose interface continuity
serves here just as well as that of ${\bf{  {B}}}(z)$ {\em{per se}}.  Thus
\begin{equation}
{\rm{\bf{\hat{e}}}}_{z}{\mbox{\boldmath$\times\,$}}{\bf{  {B}}}(z)=\left\{\rule{0mm}{1.9cm}\begin{array}{lcl}
(k_{1}/\omega)\,
\left(\rule{0mm}{5mm}e^{-ik_{1}z}{\bf{  {E}}}_{{\rm{inc}}}-e^{ik_{1}z}{\bf{  {E}}}_{{\rm{ref}}}\right) & ; & z>0 \\
&      &      \\
 i(p_{+}/\omega)\,
\left(\rule{0mm}{5mm}e^{p_{+}z}{\bf{  {E}}}_{{\rm{slab}}}^{+}-e^{-p_{+}z}{\bf{  {E}}}_{{\rm{slab}}}^{-}\right) & ; & 0>z>-a \\
&       &      \\
 (k_{1}/\omega)\,
\left(\rule{0mm}{5mm}e^{-ik_{1}z}{\bf{  {E}}}_{{\rm{trans}}}\right) & ; & -a>z                       
\end{array} \right. \,.
\end{equation}
The requisite electric interface continuity equations can now be written as
\begin{equation}
\left\{\rule{0mm}{1.2cm}\begin{array}{lcl}
{\bf{  {E}}}_{{\rm{inc}}}+{\bf{  {E}}}_{{\rm{ref}}} & = & {\bf{  {E}}}_{{\rm{slab}}}^{+}+{\bf{  {E}}}_{{\rm{slab}}}^{-} \\
           &       &     \\           
e^{-p_{+}a}{\bf{  {E}}}_{{\rm{slab}}}^{+}+e^{p_{+}a}{\bf{  {E}}}_{{\rm{slab}}}^{-} & = &
         e^{ik_{1}a}{\bf{  {E}}}_{{\rm{trans}}}
         \end{array} \right. 
\end{equation}
\newpage
\mbox{    }
\newline
\newline
\newline
and are joined as
\begin{equation}
\left\{\rule{0mm}{1.3cm}\begin{array}{lcl}
k_{1}\!
\left(\rule{0mm}{4.5mm}{\bf{  {E}}}_{{\rm{inc}}}-{\bf{  {E}}}_{{\rm{ref}}}\right)   & = &
 ip_{+}\left(\rule{0mm}{4.5mm}{\bf{  {E}}}_{{\rm{slab}}}^{+}-{\bf{  {E}}}_{{\rm{slab}}}^{-}\right) \\
             &      &      \\
ip_{+}
\left(\rule{0mm}{4.5mm}e^{-p_{+}a}{\bf{  {E}}}_{{\rm{slab}}}^{+}-e^{p_{+}a}{\bf{  {E}}}_{{\rm{slab}}}^{-}\right) & = & 
               k_{1}\left(\rule{0mm}{4.5mm}e^{ik_{1}a}{\bf{  {E}}}_{{\rm{trans}}}\right)    
                        \end{array} \right. 
\end{equation}
by their magnetic partners.

     While linear system (48)-(49) can of course be solved {\em{en masse}} for the unknowns
${\bf{  {E}}}_{{\rm{ref}}},$ ${\bf{  {E}}}_{{\rm{trans}}},$ and ${\bf{  {E}}}_{{\rm{slab}}}^{\pm}$ through recourse to a four-by-four determinant, we prefer to straddle it instead via a cascade of two-by-two linear systems.  Thus, on grouping the first lines we obtain a two-by-two system for ${\bf{  {E}}}_{{\rm{slab}}}^{\pm}$ having its source built up from ${\bf{  {E}}}_{{\rm{inc}}}$ and ${\bf{  {E}}}_{{\rm{ref}}}.$  When instead the second lines are so grouped, the source becomes controlled by ${\bf{  {E}}}_{{\rm{trans}}}.$  A final demand that the seemingly discordant solutions for ${\bf{  {E}}}_{{\rm{slab}}}^{\pm}$ so gotten be in fact identical will then emerge as a two-by-two linear system for
${\bf{  {E}}}_{{\rm{ref}}}$ and ${\bf{  {E}}}_{{\rm{trans}}}$ alone, allowing us to recover outcomes (44) and  (45) as previously encountered.

     And so, from
\begin{equation}
\left[\begin{array}{lr}
            1 &  1 \\
              &    \\
            1 & -1 
        \end{array}\right]\left[\begin{array}{c}
                  {\bf{  {E}}}_{{\rm{slab}}}^{+} \\
                                                     \\
                  {\bf{  {E}}}_{{\rm{slab}}}^{-}
                  \end{array} \right]=\left[\begin{array}{c}
                    {\bf{  {E}}}_{{\rm{inc}}}+{\bf{  {E}}}_{{\rm{ref}}} \\
                                                                            \\
  -ik_{1}/p_{+}\left(\rule{0mm}{4mm}{\bf{  {E}}}_{{\rm{inc}}}-{\bf{  {E}}}_{{\rm{ref}}}\right)
                   \end{array} \right]
\end{equation}
we obtain
\begin{equation}
\left[\begin{array}{c}
{\bf{  {E}}}_{{\rm{slab}}}^{+} \\
\\
{\bf{  {E}}}_{{\rm{slab}}}^{-}
\end{array} \right]=\frac{1}{2}\left[\begin{array}{c}
\left(\rule{0mm}{4mm}1-ik_{1}/p_{+}\right){\bf{  {E}}}_{{\rm{inc}}}+
\left(\rule{0mm}{4mm}1+ik_{1}/p_{+}\right){\bf{  {E}}}_{{\rm{ref}}} \\
                            \\
\left(\rule{0mm}{4mm}1+ik_{1}/p_{+}\right){\bf{  {E}}}_{{\rm{inc}}}+
\left(\rule{0mm}{4mm}1-ik_{1}/p_{+}\right){\bf{  {E}}}_{{\rm{ref}}}
\end{array} \right]\,,
\end{equation}
whereas
\begin{equation}
\left[\begin{array}{lr}
e^{-p_{+}a} &  e^{p_{+}a} \\
&    \\
e^{-p_{+}a} & -e^{p_{+}a} 
\end{array}\right]\left[\begin{array}{c}
{\bf{  {E}}}_{{\rm{slab}}}^{+} \\
\\
{\bf{  {E}}}_{{\rm{slab}}}^{-}
\end{array} \right]=\left[\begin{array}{c}
1 \\
\\
-ik_{1}/p_{+}
\end{array} \right]e^{ik_{1}a}{\bf{  {E}}}_{{\rm{trans}}}
\end{equation}
similarly gives                                                                             
\begin{equation}
\left[\begin{array}{c}
{\bf{  {E}}}_{{\rm{slab}}}^{+} \\
\\
{\bf{  {E}}}_{{\rm{slab}}}^{-}
\end{array} \right]=\frac{1}{2}\left[\begin{array}{lr}
e^{p_{+}a} &  e^{p_{+}a} \\
&    \\
e^{-p_{+}a} & -e^{-p_{+}a} 
\end{array}\right]\left[\begin{array}{c}
1 \\
\\
-ik_{1}/p_{+}
\end{array} \right]e^{ik_{1}a}{\bf{  {E}}}_{{\rm{trans}}}\,,
\end{equation}
or else
\begin{equation}
\left[\begin{array}{c}
{\bf{  {E}}}_{{\rm{slab}}}^{+} \\
\\
{\bf{  {E}}}_{{\rm{slab}}}^{-}
\end{array} \right]=\frac{1}{2}\left[\begin{array}{c}
\left(\rule{0mm}{4mm}1-ik_{1}/p_{+}\right)e^{p_{+}a} \\
\\
\left(\rule{0mm}{4mm}1+ik_{1}/p_{+}\right)e^{-p_{+}a}
\end{array} \right]e^{ik_{1}a}{\bf{  {E}}}_{{\rm{trans}}}\,.
\end{equation}
A final demand that (51) and (54) stand in agreement amounts to the statement that
\begin{equation}
\left[\begin{array}{lr}
  \left(\rule{0mm}{4mm}1+ik_{1}/p_{+}\right)  & - \left(\rule{0mm}{4mm}1-ik_{1}/p_{+}\right)e^{(ik_{1}+p_{+})a}  \\
                                       &        \\
  \left(\rule{0mm}{4mm}1-ik_{1}/p_{+}\right)  & - \left(\rule{0mm}{4mm}1+ik_{1}/p_{+}\right)e^{(ik_{1}-p_{+})a} 
  \end{array}\right]\left[\begin{array}{l}
  {\bf{  {E}}}_{{\rm{ref}}} \\
  \\
  {\bf{  {E}}}_{{\rm{trans}}}
  \end{array} \right]=-\left[\begin{array}{c}
              \left(\rule{0mm}{4mm}1-ik_{1}/p_{+}\right) \\
                                                         \\
             \left(\rule{0mm}{4mm}1+ik_{1}/p_{+}\right)
             \end{array}\right]{\bf{  {E}}}_{{\rm{inc}}}\,.                                                                   
 \end{equation}
  \newpage
 \mbox{    }
 \newline
 \newline
 \newline
Recalling next the existing abbreviations $-ik_{1}+p_{\pm}=\gamma_{\pm},$ (55) can be condensed into
\begin{equation}
\left[\begin{array}{rr}
-\gamma_{-}  & -\gamma_{+}e^{-\,\gamma_{-}a}  \\
&        \\
\gamma_{+}  & \gamma_{-}e^{-\,\gamma_{+}a} 
\end{array}\right]\left[\begin{array}{l}
{\bf{  {E}}}_{{\rm{ref}}} \\
\\
{\bf{  {E}}}_{{\rm{trans}}}
\end{array} \right]=\left[\begin{array}{r}
-\gamma_{+} \\
\\
\gamma_{-}
\end{array}\right]{\bf{  {E}}}_{{\rm{inc}}}\,.                                                                   
\end{equation}
whose solution
\begin{equation}
\left[\begin{array}{l}
{\bf{  {E}}}_{{\rm{ref}}} \\
\\
{\bf{  {E}}}_{{\rm{trans}}}
\end{array} \right]=\frac{ e^{\gamma_{-}a}}{\gamma_{+}^{2}-\gamma_{-}^{2}e^{(\gamma_{-}-\,\gamma_{+})a}}
\left[\begin{array}{rr}
\gamma_{-}e^{-\,\gamma_{+}a}  & \gamma_{+}e^{-\,\gamma_{-}a}  \\
&        \\
-\gamma_{+}  & -\gamma_{-} 
\end{array}\right]\left[\begin{array}{r}
-\gamma_{+} \\
\\
\gamma_{-}
\end{array}\right]{\bf{  {E}}}_{{\rm{inc}}}
\end{equation}
recovers (44) and the third line from (45) as to their every detail.  ${\bf{  {E}}}_{{\rm{slab}}}^{\pm}$ can of course
now simply be read off from either (51) or (53).  This labyrinthine odyssey is thus graced
by success and so bestows at least a partial nod of approval upon our radiative self-consistency program.
\section{Appendix:  radiative self-consistency with dissimilar half spaces}
    Should we allow the lower half space to differ from its upper counterpart, having now a dielectric constant
$\epsilon_{3}\neq\epsilon_{1}$ and arbitrary conductivity $\sigma_{3},$ it being understood however that at least one of
$\epsilon_{3}$ and $\sigma_{3}$ differs respectively from $\epsilon_{2}$ and $\sigma_{2},$ then a corresponding provision
would have to be made for additional radiative sources seen, as before, from the perspective of the nondissipative medium 1.

     The total electric field in the lower half space can now be taken as
     \begin{eqnarray}
     {\bf{  {E}}}_{{\rm{tot}}}({\bf{r}}) & = & {\bf{  {E}}}_{{\rm{half-low}}}^{+}
     \exp\left(\rule{0mm}{1.5mm}ik_{1}{\rm{\bf{\hat{n}}}}_{\|}
     {\mbox{\boldmath$\cdot$}}{\mbox{\boldmath$\rho$}}+ q_{+}z\right)
     \end{eqnarray}
     with
     \begin{equation}
     q_{+}^{2}
     -k_{1}^{2}{\rm{\bf{\hat{n}}}}_{\|}{\mbox{\boldmath$\cdot\,$}}{\rm{\bf{\hat{n}}}}_{\|}=
     -\omega^{2}\mu\epsilon_{3}\left(\rule{0mm}{4mm}1+i\sigma_{3}/\omega\epsilon_{3}\right)
     \end{equation}
     and hence
     \begin{eqnarray}
     q_{+} &  = &   k_{1}\sqrt{\rule{0mm}{4mm}\sin^{2}\vartheta-\left(\rule{0mm}{4mm}\epsilon_{3}/\epsilon_{1}\right)
     	\!\left(\rule{0mm}{4mm}1+i\sigma_{3}/\omega\epsilon_{3}\right)\,} \nonumber  \\
     & = & \frac{k_{1}}{\sqrt{2\,}} \left\{\rule{0mm}{8mm}
     \sqrt{\sqrt{\rule{0mm}{4mm}\left(\rule{0mm}{4mm}\sin^{2}\vartheta-\epsilon_{3}/\epsilon_{1}\right)^{2}+
     		\left(\rule{0mm}{4mm}\sigma_{3}/\omega\epsilon_{1}\right)^{2}\,}+\sin^{2}\vartheta-\epsilon_{3}/\epsilon_{1}\,}  \right.  \\
     &    &  \rule{-0.3cm}{0mm} \left.\rule{0mm}{8mm} -i\frac{\omega}{|\,\omega\,|}
     \sqrt{\sqrt{\rule{0mm}{4mm}\left(\rule{0mm}{4mm}\sin^{2}\vartheta-\epsilon_{3}/\epsilon_{1}\right)^{2}+
     		\left(\rule{0mm}{4mm}\sigma_{3}/\omega\epsilon_{1}\right)^{2}\,}-\sin^{2}\vartheta+\epsilon_{3}/\epsilon_{1}\,}  \;\right\}\,.\nonumber			
     \end{eqnarray}
     Moreover, if one sets
     \begin{equation}
     {\bf{   {Q}}}^{+}=ik_{1}{\rm{\bf{\hat{n}}}}_{\|}+q_{+}{\rm{\bf{\hat{e}}}}_{z}\,,
     \end{equation}
     \newpage
     \mbox{    }
     \newline
     \newline
     \newline
     then the requisite divergenceless nature of field (58), seen now in the form
     ${\bf{  {E}}}_{{\rm{half-low}}}^{+}
     \exp\left(\rule{0mm}{3.5mm}{\bf{   {Q}}^{+}}{\mbox{\boldmath$\cdot\,$}}{\rm{\bf{r}}}\right),$ is assured by having 
     \begin{equation}
     {\bf{   {Q}}}^{+}{\mbox{\boldmath$\cdot\,$}}{\bf{  {E}}}_{{\rm{half-low}}}^{+}=0\,.
     \end{equation}     
As the counterpart to (7), the total magnetic field now becomes
\begin{eqnarray}
\lefteqn{
{\bf{  {B}}}_{{\rm{tot}}}({\bf{r}}) =\frac{k_{1}}{\omega}
{\rm{\bf{\hat{n}}}}{\mbox{\boldmath$\,\times$}}{\bf{  {E}}}_{{\rm{inc}}}
e^{ik_{1}{\rm{\bf{\hat{n}}}}{\mbox{\boldmath$\cdot$}}{\bf{r}}} } \nonumber  \\
  &  & \rule{-0.1cm}{0mm} \frac{\mu(\sigma_{2}-i\omega\{\epsilon_{2}-\epsilon_{1}\})}{4\pi}
\sum_{\pm} 
{\mbox{\boldmath$\nabla\times$}}\left\{\rule{0mm}{7mm}{\bf{  {E}}}_{{\rm{slab}}}^{\pm}
\int_{0>z'>-a}\frac{e^{ik_{1}|{\bf{r}}-{\bf{r'}}|}}{|{\bf{r}}-{\bf{r'}}|}     
\exp\left(\rule{0mm}{1.5mm}ik_{1}{\rm{\bf{\hat{n}}}}_{\|}
{\mbox{\boldmath$\cdot$}}{\mbox{\boldmath${\tiny{\rho'}}$}}+ p_{\pm}z'\right)dx'dy'dz'\right\}  \nonumber \\
    &   & \rule{-0.5cm}{0mm} + \frac{\mu(\sigma_{3}-i\omega\{\epsilon_{3}-\epsilon_{1}\})}{4\pi}    
    {\mbox{\boldmath$\nabla\times$}}\left\{\rule{0mm}{7mm}{\bf{  {E}}}_{{\rm{half-low}}}^{+}
    \int_{-a>z'}\frac{e^{ik_{1}|{\bf{r}}-{\bf{r'}}|}}{|{\bf{r}}-{\bf{r'}}|}     
    \exp\left(\rule{0mm}{1.5mm}ik_{1}{\rm{\bf{\hat{n}}}}_{\|}
    {\mbox{\boldmath$\cdot$}}{\mbox{\boldmath${\tiny{\rho'}}$}}+ q_{+}z'\right)dx'dy'dz'\right\}
\end{eqnarray}           
and leads, just a before, to the revised integral equation 
\begin{eqnarray}
\lefteqn{
	{\hat{\epsilon}}(z){\bf{  {E}}}_{{\rm{tot}}}({\bf{r}})  =  
	\epsilon_{1}{\bf{  {E}}}_{{\rm{inc}}}e^{ik_{1}{\rm{\bf{\hat{n}}}}{\mbox{\boldmath$\cdot$}}{\bf{r}}} } \nonumber \\
& & +\frac{i(\sigma_{2}-i\omega\{\epsilon_{2}-\epsilon_{1}\})}{4\pi\omega}
\sum_{\pm}{\mbox{\boldmath$\nabla\times$}}\left[\rule{0mm}{7mm} 
{\mbox{\boldmath$\,\nabla\times$}}\left\{\rule{0mm}{7mm}	{\bf{  {E}}}_{{\rm{slab}}}^{\pm}
\int_{0>z'>-a}\frac{e^{ik_{1}|{\bf{r}}-{\bf{r'}}|}}{|{\bf{r}}-{\bf{r'}}|}     
\exp\left(\rule{0mm}{1.5mm}ik_{1}{\rm{\bf{\hat{n}}}}_{\|}
{\mbox{\boldmath$\cdot$}}{\mbox{\boldmath$\rho'$}}+ p_{\pm}z'\right)dx'dy'dz'\right\}\,\right] \rule{0mm}{7mm} \nonumber \\
 & &  +\frac{i(\sigma_{3}-i\omega\{\epsilon_{3}-\epsilon_{1}\})}{4\pi\omega}
 {\mbox{\boldmath$\nabla\times$}}\left[\rule{0mm}{7mm} 
 {\mbox{\boldmath$\,\nabla\times$}}\left\{\rule{0mm}{7mm}	{\bf{  {E}}}_{{\rm{half-low}}}^{+}
 \int_{-a>z'}\frac{e^{ik_{1}|{\bf{r}}-{\bf{r'}}|}}{|{\bf{r}}-{\bf{r'}}|}     
 \exp\left(\rule{0mm}{1.5mm}ik_{1}{\rm{\bf{\hat{n}}}}_{\|}
 {\mbox{\boldmath$\cdot$}}{\mbox{\boldmath$\rho'$}}+ q_{+}z'\right)dx'dy'dz'\right\}\,\right] 
\end{eqnarray}
valid everywhere without exception, provided only that we modify the last entry in (8) so as to read
${\hat{\epsilon}}(z)=\epsilon_{3}+i\sigma_{3}/\omega$ whenever $-a>z,$

     The half space source additions to (63)-(64) entail a fresh calculation
\begin{eqnarray}
\int_{-a>z'}\frac{e^{ik_{1}|{\bf{r}}-{\bf{r'}}|}}{|{\bf{r}}-{\bf{r'}}|}     
\exp\left(ik_{1}{\rm{\bf{\hat{n}}}}_{\|}
{\mbox{\boldmath$\cdot$}}{\mbox{\boldmath$\rho'$}}+ q_{+}z'\right)dx'dy'dz' & = &     \nonumber  \\
&  & \rule{-6.6cm}{0mm}=\,2\pi \exp\left(ik_{1}{\rm{\bf{\hat{n}}}}_{\|}
{\mbox{\boldmath$\cdot$}}{\mbox{\boldmath$\rho$}}\right)\int_{-\infty}^{-a}dz'\exp\left(\rule{0mm}{3.5mm}q_{+}z'\right)
\int_{0}^{\,\infty}
\frac{e^{ik_{1}\sqrt{\rule{0mm}{3mm}\rho\,'^{\,2}+(z-z')^{2}\,}}}{\sqrt{\rule{0mm}{3mm}\rho\,'^{\,2}+(z-z')^{2}\,}}\,  
\rho'J_{0}(k_{1}\rho'\sin\vartheta)d\rho'   \nonumber   \\
&   & \rule{-6.6cm}{0mm}=\, \frac{2\pi}{ik_{1}{\hat{n}}_{z}}\exp\left(ik_{1}{\rm{\bf{\hat{n}}}}_{\|}
{\mbox{\boldmath$\cdot$}}{\mbox{\boldmath$\rho$}}\right)\int_{-\infty}^{-a}
\exp\left(\rule{0mm}{3.5mm}q_{+}z'-ik_{1}{\hat{n}}_{z}|z-z'|\right)dz'   
\end{eqnarray}
whose final integral over $z'$ must be separately considered for $z>-a$ (values $K_{1}(z)$) and $-a>z$ (values
$K_{2}(z)$).  We find:

\parindent=0in
{\bf{\underline{{\mbox{\boldmath$K_{1}(z),\; z\!>\!-a$}}}:}}         
\parindent=0.5in     
\begin{eqnarray}
K_{1}(z) & = & \exp\left(\rule{0mm}{3.5mm}\!-ik_{1}{\hat{n}}_{z}z\right)\int_{-\infty}^{-a}
\exp\left(\!\rule{0mm}{4mm}\left\{\rule{0mm}{3.5mm}ik_{1}{\hat{n}}_{z}+q_{+}\!\right\}\!z'\right)dz'  \nonumber \\
& = & \exp\left(\rule{0mm}{3.5mm}\!-ik_{1}{\hat{n}}_{z}z\right)
\left\{\rule{0mm}{3.5mm}ik_{1}{\hat{n}}_{z}+q_{+}\!\right\}^{-1}
\exp\left(\rule{0mm}{4mm}-\left\{\rule{0mm}{3.5mm}ik_{1}{\hat{n}}_{z}+q_{+}\!\right\}a\right)\,;  
\end{eqnarray}
\newpage
\mbox{    }
\newline
\newline
\newline
and

\parindent=0in
{\bf{\underline{{\mbox{\boldmath$K_{2}(z),\; -a\!>\!z$}}}:}}         
\parindent=0.5in 
\begin{eqnarray}
K_{2}(z) & = & \exp\left(\rule{0mm}{3.5mm}\!-ik_{1}{\hat{n}}_{z}z\right)\int_{-\infty}^{\,z}
\exp\left(\!\rule{0mm}{4mm}\left\{\rule{0mm}{3.5mm}ik_{1}{\hat{n}}_{z}+q_{+}\!\right\}\!z'\right)dz'  \nonumber \\
 & & + \exp\left(\rule{0mm}{3.5mm}ik_{1}{\hat{n}}_{z}z\right)\int_{\,z}^{-a}
 \exp\left(\!\rule{0mm}{4mm}-\left\{\rule{0mm}{3.5mm}ik_{1}{\hat{n}}_{z}-q_{+}\!\right\}\!z'\right)dz'       \\
 & = & \left\{\frac{1}{ik_{1}{\hat{n}}_{z}+q_{+}}+\frac{1}{ik_{1}{\hat{n}}_{z}-q_{+}}\right\}e^{q_{+}z}-
 \frac{e^{\{ik_{1}{\hat{n}}_{z}-q_{+}\}a}}{ik_{1}{\hat{n}}_{z}-q_{+}}e^{ik_{1}{\hat{n}}_{z}z}\,.  \nonumber
\end{eqnarray}

    From (64)-(66) we now find that the total field ${\bf{  {E}}}_{{\rm{tot}}}({\bf{r}})$ as first encountered in
(17) for $z>0,$ must be augmented so as to read
\begin{eqnarray}
\lefteqn{
\rule{-5mm}{0mm}{\bf{  {E}}}_{{\rm{tot}}}({\bf{r}}) = 
{\bf{  {E}}}_{{\rm{inc}}}e^{ik_{1}{\rm{\bf{\hat{n}}}}{\mbox{\boldmath$\cdot$}}{\bf{r}}} } \nonumber \\
 &   &  -\frac{k_{1}(\sigma_{2}-i\omega\{\epsilon_{2}-\epsilon_{1}\})}{2{\hat{n}}_{z}\epsilon_{1}\omega}\,
e^{ik_{1}{\rm{\bf{\hat{n}}'}}{\mbox{\boldmath$\cdot\,$}}{\bf{r}}}\,
{\rm{\bf{\hat{n}}'}}{\mbox{\boldmath$\,\times$}}
\left[\rule{0mm}{5mm}\,{\rm{\bf{\hat{n}}'}}{\mbox{\boldmath$\,\times$}}\sum_{\pm}\left\{\frac{1-
	\exp\left(\!\rule{0mm}{4mm}-\left\{\rule{0mm}{3.5mm}ik_{1}{\hat{n}}_{z}+p_{\pm}\right\}\!a\right)}
{ik_{1}{\hat{n}}_{z}+p_{\pm}}\right\}{\bf{  {E}}}_{{\rm{slab}}}^{\pm}\,\right]    \\
 &   &  -\frac{k_{1}(\sigma_{3}-i\omega\{\epsilon_{3}-\epsilon_{1}\})}{2{\hat{n}}_{z}\epsilon_{1}\omega}\,
e^{ik_{1}{\rm{\bf{\hat{n}}'}}{\mbox{\boldmath$\cdot\,$}}{\bf{r}}}\,
{\rm{\bf{\hat{n}}'}}{\mbox{\boldmath$\,\times$}}
\left[\rule{0mm}{7mm}\,{\rm{\bf{\hat{n}}'}}{\mbox{\boldmath$\,\times$}}
\left\{\rule{0mm}{6mm}
\frac{\exp\left(\rule{0mm}{4mm}-\left\{\rule{0mm}{3.5mm}ik_{1}{\hat{n}}_{z}+q_{+}\!\right\}a\right)}	
{ik_{1}{\hat{n}}_{z}+q_{+}}\right\}{\bf{  {E}}}_{{\rm{half-low}}}^{+}\,\right] \,.  \nonumber
\end{eqnarray}

      Turning attention with the aid of (67) to the lower half space as such, we compute, on the basis of (60), that
\begin{eqnarray}
\left\{\frac{1}{ik_{1}{\hat{n}}_{z}+q_{+}}+\frac{1}{ik_{1}{\hat{n}}_{z}-q_{+}}\right\}=
-\frac{2ik_{1}{\hat{n}}_{z}}{\,k_{1}^{2}{\hat{n}}_{z}^{2}+q_{+}^{2}\,}=
\frac{2k_{1}{\hat{n}}_{z}}{\omega\mu\left(\rule{0mm}{4mm}\sigma_{3}-i\omega\left\{\epsilon_{3}-\epsilon_{1}\right\}\right)}\,,
\end{eqnarray}
akin to what had been earlier found in (25).  The associated cluster from (67) thus contributes
\begin{eqnarray}
\left(\rule{0mm}{4mm}\epsilon_{3}+i\sigma_{3}/\omega\right)
{\bf{  {E}}}_{{\rm{half-low}}}^{+}
\exp\left(\rule{0mm}{3.5mm}{\bf{   {Q}}^{+}}{\mbox{\boldmath$\cdot\,$}}{\rm{\bf{r}}}\right)
\end{eqnarray}
to the third line of integral equation (64), a contribution which, in the distant spirit of Ewald-Oseen ({\em{cf.}} the
closing comments in Note 5 at paper's end), cancels
exactly once again the term ${\hat{\epsilon}}(z){\bf{  {E}}}_{{\rm{tot}}}({\bf{r}})$ on its left.  Orthogonality
(62) participates decisively in the intervening algebra, as do Eqs. (59) and (65).

      The second cluster in the third line from (67) produces by contrast a contribution whose phase progression
$e^{ik_{1}{\rm{\bf{\hat{n}}}}{\mbox{\boldmath$\cdot$}}{\bf{r}}}$ mimics that of the incoming excitation.  After
some algebra we find this contribution to be
\begin{eqnarray}
  &   & \frac{k_{1}(\sigma_{3}-i\omega\{\epsilon_{3}-\epsilon_{1}\})}{2{\hat{n}}_{z}\omega}\,
     e^{ik_{1}{\rm{\bf{\hat{n}}}}{\mbox{\boldmath$\cdot$}}{\bf{r}}}\,
     {\rm{\bf{\hat{n}}}}{\mbox{\boldmath$\,\times$}}
     \left[\rule{0mm}{7mm}\,{\rm{\bf{\hat{n}}}}{\mbox{\boldmath$\,\times$}}\left\{\rule{0mm}{6mm}
     \frac{e^{\{ik_{1}{\hat{n}}_{z}-q_{+}\}a}}	
     {ik_{1}{\hat{n}}_{z}-q_{+}}\right\}{\bf{  {E}}}_{{\rm{half-low}}}^{+}\,\right].   
\end{eqnarray}
\newpage
\mbox{   }
\newline
\newline
\newline
This latter must of course be compounded with $\epsilon_{1}$ times the transmitted structure from (18) so as to yield
a net result of ${\bf{0}}.$  Taking note of (22), which retains validity unimpaired, and the abbreviations introduced
under (24)-(25), $\epsilon_{1}$ times the content of (18) on its right can be rendered as
\begin{eqnarray}
\rule{3mm}{0mm}
    &  & \frac{ik_{1}}{2{\hat{n}}_{z}\omega^{2}\mu}\,
 e^{ik_{1}{\rm{\bf{\hat{n}}}}{\mbox{\boldmath$\cdot$}}{\bf{r}}}
{\rm{\bf{\hat{n}}}}{\mbox{\boldmath$\,\times$}}\!
\left[\rule{0mm}{6mm}\,{\rm{\bf{\hat{n}}}}{\mbox{\boldmath$\,\times\,$}}\!\sum_{\pm}
	\gamma_{\pm}e^{\gamma_{\mp}a}{\bf{  {E}}}_{{\rm{slab}}}^{\pm}\,\right] 
\end{eqnarray}
and hence leads us to demand that
\begin{eqnarray}
 {\rm{\bf{\hat{n}}}}{\mbox{\boldmath$\,\times$}}\!
  \left[\rule{0mm}{6mm}\,{\rm{\bf{\hat{n}}}}{\mbox{\boldmath$\,\times\,$}}\!\sum_{\pm}
  \gamma_{\pm}e^{\gamma_{\mp}a}{\bf{  {E}}}_{{\rm{slab}}}^{\pm}\,\right] 
      & = &    i\omega\mu\left(\rule{0mm}{4mm}\sigma_{3}-i\omega\{\epsilon_{3}-\epsilon_{1}\}\right)\,
{\rm{\bf{\hat{n}}}}{\mbox{\boldmath$\,\times$}}
\left[\rule{0mm}{7mm}\,{\rm{\bf{\hat{n}}}}{\mbox{\boldmath$\,\times$}}\left\{\rule{0mm}{6mm}
\frac{e^{\{ik_{1}{\hat{n}}_{z}-q_{+}\}a}}	
{ik_{1}{\hat{n}}_{z}-q_{+}}\right\}{\bf{  {E}}}_{{\rm{half-low}}}^{+}\,\right]
\end{eqnarray}                
as the first amplitude condition which brings in the participation of ${\bf{  {E}}}_{{\rm{half-low}}}^{+}.$ 

     Yet another appearance of ${\bf{  {E}}}_{{\rm{half-low}}}^{+}$ arises by virtue of cross-talk in reverse
between the lower half space and the slab as expressed by the third line from (68), the structure
of that line being uniformly applicable whenever $z>-a,$ and not merely $z>0$ when subordinated as a
contributor to (68).   While (22) remains unaffected, reference to (21) shows that (23) now blends into
\begin{eqnarray}
\lefteqn{
{\rm{\bf{\hat{n}'}}}{\mbox{\boldmath$\,\times$}}\!
\left[\rule{0mm}{7mm}\,{\rm{\bf{\hat{n}'}}}{\mbox{\boldmath$\,\times\,$}}\!\sum_{\pm}
\gamma_{\mp}e^{-\,\gamma_{\pm}a}{\bf{  {E}}}_{{\rm{slab}}}^{\pm}\,\right]  =  }      \\ 
 & \rule{1.5cm}{0mm} = &   i\omega\mu\left(\rule{0mm}{4mm}\sigma_{3}-i\omega\{\epsilon_{3}-\epsilon_{1}\}\right)\,      
      {\rm{\bf{\hat{n}}'}}{\mbox{\boldmath$\,\times$}}
      \left[\rule{0mm}{7mm}\,{\rm{\bf{\hat{n}}'}}{\mbox{\boldmath$\,\times$}}\left\{\rule{0mm}{6mm}
 \frac{\exp\left(\rule{0mm}{4mm}-\left\{\rule{0mm}{3.5mm}ik_{1}{\hat{n}}_{z}+q_{+}\!\right\}a\right)}	
      {ik_{1}{\hat{n}}_{z}+q_{+}}\right\}{\bf{  {E}}}_{{\rm{half-low}}}^{+}\,\right].  \nonumber  
\end{eqnarray}

     In principle, the triplet of vector equations (22), (73)-(74), nine in number, component by component,
is just adequate to determine in full, via routine linear algebra methods, the vector amplitudes
${\bf{  {E}}}_{{\rm{slab}}}^{\pm}$ and ${\bf{  {E}}}_{{\rm{half-low}}}^{+}.$  Nevertheless we would like
to indicate here at least the first steps
of an invariantive, vector solution path, resembling that previously elaborated in Eq. (26) onward.
As before we can ease the notation somewhat by writing $\zeta_{\pm}=ik_{1}{\hat{n}}_{z}\pm q_{+},$ so that
$\zeta_{\pm}\zeta_{\mp}=i\omega\mu\left(\rule{0mm}{4mm}\sigma_{3}-i\omega\{\epsilon_{3}-\epsilon_{1}\}\right).$
This allows (73)-(74) to be restated more concisely as
\begin{eqnarray}
{\rm{\bf{\hat{n}}}}{\mbox{\boldmath$\,\times$}}\!
\left[\rule{0mm}{6mm}\,{\rm{\bf{\hat{n}}}}{\mbox{\boldmath$\,\times\,$}}\!\sum_{\pm}
\gamma_{\pm}e^{\gamma_{\mp}a}{\bf{  {E}}}_{{\rm{slab}}}^{\pm}\,\right] 
& = &    \zeta_{+}e^{\zeta_{-}a}\,
{\rm{\bf{\hat{n}}}}{\mbox{\boldmath$\,\times$}}
\left(\rule{0mm}{5mm}\,{\rm{\bf{\hat{n}}}}{\mbox{\boldmath$\,\times\,$}}{\bf{  {E}}}_{{\rm{half-low}}}^{+}\,\right)
\end{eqnarray}
and
\begin{eqnarray}
{\rm{\bf{\hat{n}'}}}{\mbox{\boldmath$\,\times$}}\!
\left[\rule{0mm}{6mm}\,{\rm{\bf{\hat{n}'}}}{\mbox{\boldmath$\,\times\,$}}\!\sum_{\pm}
\gamma_{\mp}e^{-\,\gamma_{\pm}a}{\bf{  {E}}}_{{\rm{slab}}}^{\pm}\,\right] & = & 
\zeta_{-}e^{-\zeta_{+}a}\,      
{\rm{\bf{\hat{n}}'}}{\mbox{\boldmath$\,\times$}}
\left(\rule{0mm}{5mm}\,{\rm{\bf{\hat{n}}'}}{\mbox{\boldmath$\,\times\,$}}{\bf{  {E}}}_{{\rm{half-low}}}^{+}\,\right).    
\end{eqnarray}
Since, say
\begin{equation}
{\bf{  {E}}}_{{\rm{half-low}}}^{+} ={\rm{\bf{\hat{n}}}}\left(\rule{0mm}{5mm}{\rm{\bf{\hat{n}}}}
{\mbox{\boldmath$\cdot\,$}}{\bf{  {E}}}_{{\rm{half-low}}}^{+}\right)-
{\rm{\bf{\hat{n}}}}{\mbox{\boldmath$\,\times$}}
\left(\rule{0mm}{5mm}\,{\rm{\bf{\hat{n}}}}{\mbox{\boldmath$\,\times\,$}}{\bf{  {E}}}_{{\rm{half-low}}}^{+}\,\right),
\end{equation}
it follows from (75) that
\begin{equation}
{\rm{\bf{\hat{n}}}}{\mbox{\boldmath$\,\cdot\,$}}{\bf{  {E}}}_{{\rm{half-low}}}^{+} =
e^{-\zeta_{-}a}\frac{\left(\rule{0mm}{4.5mm}{\bf{   {Q}}^{+}}{\mbox{\boldmath$\cdot\,$}}
	\left\{\rule{0mm}{4.5mm}{\rm{\bf{\hat{n}}}}{\mbox{\boldmath$\,\times$}}\!
	\left[\rule{0mm}{4.5mm}\,{\rm{\bf{\hat{n}}}}{\mbox{\boldmath$\,\times\,$}}\!\sum_{\pm}
	\gamma_{\pm}e^{\gamma_{\mp}a}{\bf{  {E}}}_{{\rm{slab}}}^{\pm}\,\right]\right\}\right)}
{\zeta_{+}\left(\rule{0mm}{3.5mm}{\bf{   {Q}}^{+}}{\mbox{\boldmath$\cdot\,$}}{\rm{\bf{\hat{n}}}}\right)}
\end{equation}
\newpage
\mbox{   }
\newline
\newline
\newline 
and then
{\small{
\begin{eqnarray}
\rule{-5mm}{0mm}
{\bf{  {E}}}_{{\rm{half-low}}}^{+} & = & \zeta_{+}^{-1}e^{-\zeta_{-}a}\left[\rule{0mm}{11mm}\,
{\rm{\bf{\hat{n}}}}\left\{\rule{0mm}{7mm}\frac{\left(\rule{0mm}{4.5mm}{\bf{   {Q}}^{+}}{\mbox{\boldmath$\cdot\,$}}
	\left\{\rule{0mm}{4.5mm}{\rm{\bf{\hat{n}}}}{\mbox{\boldmath$\,\times$}}\!
	\left[\rule{0mm}{4.5mm}\,{\rm{\bf{\hat{n}}}}{\mbox{\boldmath$\,\times\,$}}\!\sum_{\pm}
	\gamma_{\pm}e^{\gamma_{\mp}a}{\bf{  {E}}}_{{\rm{slab}}}^{\pm}\,\right]\right\}\right)}
{\left(\rule{0mm}{3.5mm}{\bf{   {Q}}^{+}}{\mbox{\boldmath$\cdot\,$}}{\rm{\bf{\hat{n}}}}\right)}\right\}
    -{\rm{\bf{\hat{n}}}}{\mbox{\boldmath$\,\times$}}\!
\left\{\rule{0mm}{6mm}\,{\rm{\bf{\hat{n}}}}{\mbox{\boldmath$\,\times\,$}}\!\sum_{\pm}
\gamma_{\pm}e^{\gamma_{\mp}a}{\bf{  {E}}}_{{\rm{slab}}}^{\pm}\,\right\}\right].  
\end{eqnarray} }  }
\vspace{-3mm}

\parindent=0.0in
This realization can then be introduced on the right in (76) and the latter, thus modified, solved in conjunction with (22) for
${\bf{  {E}}}_{{\rm{slab}}}^{\pm}$ by a suitably generalized variant of our previous method from Sections 7 and 8.  But, as
the algebra threatens by now to burst its seams, we pause here in midstream, with the hope of perhaps finishing this exercise at some later date.

\begingroup
\parindent 0pt
\parskip 2ex
\def\enotesize{\normalsize}                 
\theendnotes 
\endgroup


\end{document}